\documentclass[a4paper,11pt]{article}
\usepackage{jinstpub} 

\usepackage[T1]{fontenc}
\usepackage[utf8]{inputenc}
\usepackage[english]{babel}
\usepackage{comment}

\usepackage{amssymb,amsfonts}
\usepackage{mathtools}
\usepackage{bm}
\usepackage{siunitx}
\usepackage[version=4]{mhchem}
\usepackage{graphicx}
\usepackage{subcaption}
\usepackage{makecell} 
\usepackage{tabularx} 
\usepackage{microtype}
\usepackage{lineno}
\usepackage{float}
\usepackage[numbers,sort&compress]{natbib}

\title{Optical characterization of microcellular PET and crystallized PEN reflectors in liquid argon for LEGEND-1000}

\author[a]{Marta Babicz,}
\author[a]{Laura Baudis,}
\author[b]{Brennan Hackett,}
\author[c]{Andreas Leonhardt,}
\author[a]{Francesco Piastra,}
\author[c]{Lorenzo Pietropaoli,}
\author[c]{Stefan Schönert,}
\author[a,1]{Gloria Senatore,\note{Corresponding author.}}
\author[a]{Esteban Tapia Peñas,}
\author[a]{and Tim van de Vendel\footnote{Now at Van Swinderen Institute for Particle Physics and Gravity, University of Groningen, 9747 AG Groningen, the Netherlands}}

\affiliation[a]{Department of Physics, Physik-Institut, Universität Zürich, 8057 Zurich, Switzerland}
\affiliation[b]{Oak Ridge National Laboratory, Oak Ridge, TN 37830, USA}
\affiliation[c]{Physik Department, Technische Universität München, 85748 Munich, Germany}
\emailAdd{gloria.senatore@physik.uzh.ch}

\abstract{
Efficient collection of liquid argon (LAr) scintillation light is essential for background rejection in the LEGEND experiment. We investigate microcellular polyethylene terephthalate (PET) and crystallized polyethylene naphthalate (PEN) as candidate reflector materials for LEGEND-1000. Their optical responses to $450$\,nm light and $128$\,nm LAr scintillation are measured in LAr and complemented by room-temperature reflectance measurements in air. Under scintillation illumination, PET coupled to a PEN foil and crystallized PEN increase the detected photoelectron yield by factors of $2.36\pm0.11$ and $1.75\pm0.10$, respectively, relative to a black-metal-velvet reference. A Geant4-based optical model yields reflectivity estimates at $450~\mathrm{nm}$ in LAr of $(90\pm4)\%$ for PET and $(81\pm5)\%$ for crystallized PEN, compared with $(80\pm5)\%$ for the Tetratex reference. These results motivate further investigation of both materials for LAr instrumentation, including their radiopurity, long-term cryogenic stability, and performance under matched detector conditions.}


\begin{document}
\maketitle
\flushbottom

\section{Introduction}

The Large Enriched Germanium Experiment for Neutrinoless Double-Beta Decay (LEGEND) searches for neutrinoless double-beta ($0\nu\beta\beta$) decay of \ce{^{76}Ge}, with $Q_{\beta\beta}\simeq2039~\mathrm{keV}$, using enriched high-purity germanium (HPGe) detectors immersed in liquid argon (LAr). The program comprises LEGEND-200, currently taking data at the Laboratori Nazionali del Gran Sasso in Italy \cite{firstresultsL200}, and LEGEND-1000, in the technical design phase \cite{legend1000preconceptualdesignreport}.

Energy depositions in LAr produce vacuum ultraviolet (VUV) scintillation light at approximately $128~\mathrm{nm}$. Detecting this light in coincidence with an HPGe signal allows events with energy deposited in the surrounding LAr to be rejected. A $0\nu\beta\beta$ decay, by contrast, is expected to deposit its energy predominantly in a single HPGe crystal. Measurements with a \ce{^{226}Ra} source illustrate the effectiveness of this technique: the survival probabilities after the LAr cut in the interval $1950$--$2090~\mathrm{keV}$ were $10.4\%$ during LEGEND-200 commissioning with $60~\mathrm{kg}$ of HPGe detectors and $30.4\%$ in its predecessor GERDA \cite{GERDA2020,Krausedissertation}.

LEGEND-200 enhances light collection with a wavelength-shifting reflector (WLSR): a copper cylinder lined with Tetratex$^\circledR$ (TTX), an expanded polytetrafluoroethylene (PTFE) film coated with tetraphenylbutadiene (TPB) by in-situ evaporation. TPB converts VUV photons into blue light, with an estimated wavelength-shifting quantum efficiency (QE) of $85\%$ in LAr \cite{GabrielaArticle}. The shifted light is reflected back towards the LAr instrumentation, which consists of wavelength-shifting fiber shrouds coupled to silicon photomultipliers. Thereby, the light collection and background suppression efficiency are increased.

LEGEND-1000 foresees an electroformed copper reentrant tube separating an inner volume of underground LAr, with reduced \ce{^{42}Ar} content, from an outer volume of atmospheric LAr containing instrumented neutron moderators. Both volumes will have active veto systems with extensive WLSR coverage \cite{legend1000preconceptualdesignreport}. The larger geometry, challenges of scaling TPB evaporation, and stringent radiopurity requirements motivate alternative reflector materials.

Here we investigate microcellular polyethylene terephthalate (PET), a diffuse reflector that can be coupled to a wavelength shifter, and crystallized polyethylene naphthalate (PEN), which combines reflection and wavelength shifting. Following the material descriptions in Section~\ref{Investigated_materials}, we present measurements in LAr and air in Sections~\ref{Characterization in LAr} and~\ref{Characterization in air}. A Geant4 model is used to infer the reflectivity at $450~\mathrm{nm}$ in LAr (Section~\ref{Monte_Carlo_simulation}), followed by a discussion of implications for detector instrumentation.

\section{Investigated materials}
\label{Investigated_materials}

The selected materials were screened by visual inspection and handling tests, including immersion in liquid nitrogen at $77~\mathrm{K}$ to assess shrinkage and opacity at cryogenic temperature. Both appeared more mechanically robust than TTX during handling; shrinkage was not quantified. Other porous sintered and expanded PTFE samples showed substantial shrinkage and were not studied further. Enhanced specular reflector (ESR), a multilayer composite of acrylic and PEN used in LAr-based experiments such as DarkSide-20k \cite{Thorpe2023DarkSide20k} and DUNE \cite{Bertolini2026XARAPUCA}, was not included in the samples measured in this work. PEN is used for HPGe detector holders in LEGEND-200 \cite{firstresultsL200} and as an inner-vessel coating in DarkSide-20k \cite{Thorpe2023DarkSide20k}. In this study, we investigate the optical response of crystallized PEN. We provide here details on the selected materials:

\begin{itemize}
     \item Microcellular PET (hereafter PET), grade M4, developed by Furukawa \cite{furukawa_mcpet}. The $0.5$ mm-thick foil has a microcellular structure and surface finish designed for high, predominantly diffuse visible reflectance.
     
    \item Crystallized PEN, produced at Oak Ridge National Laboratory by 3D printing with filament from FLXR Engineering \cite{flxr_pennatural}. The $1.7$ mm-thick cylinder was sized to fit the LArS sample holder and annealed at the Technical University of Munich at $200~^\circ\mathrm{C}$ for $60~\mathrm{min}$ in a nitrogen atmosphere, followed by slow cooling in air. The material became white after annealing; its crystallinity was not quantified. We retain the term ``crystallized PEN'' for this sample throughout the text. Previous measurements of PEN in LAr yielded an expected QE of $69\%$ \cite{GabrielaArticle}; this value was not independently measured for the printed cylinder.
\end{itemize}

\begin{figure}[htbp]
    \centering

    \begin{minipage}[c]{0.43\textwidth}
        \vspace{0pt}
        \centering
        \includegraphics[width=0.8\linewidth]{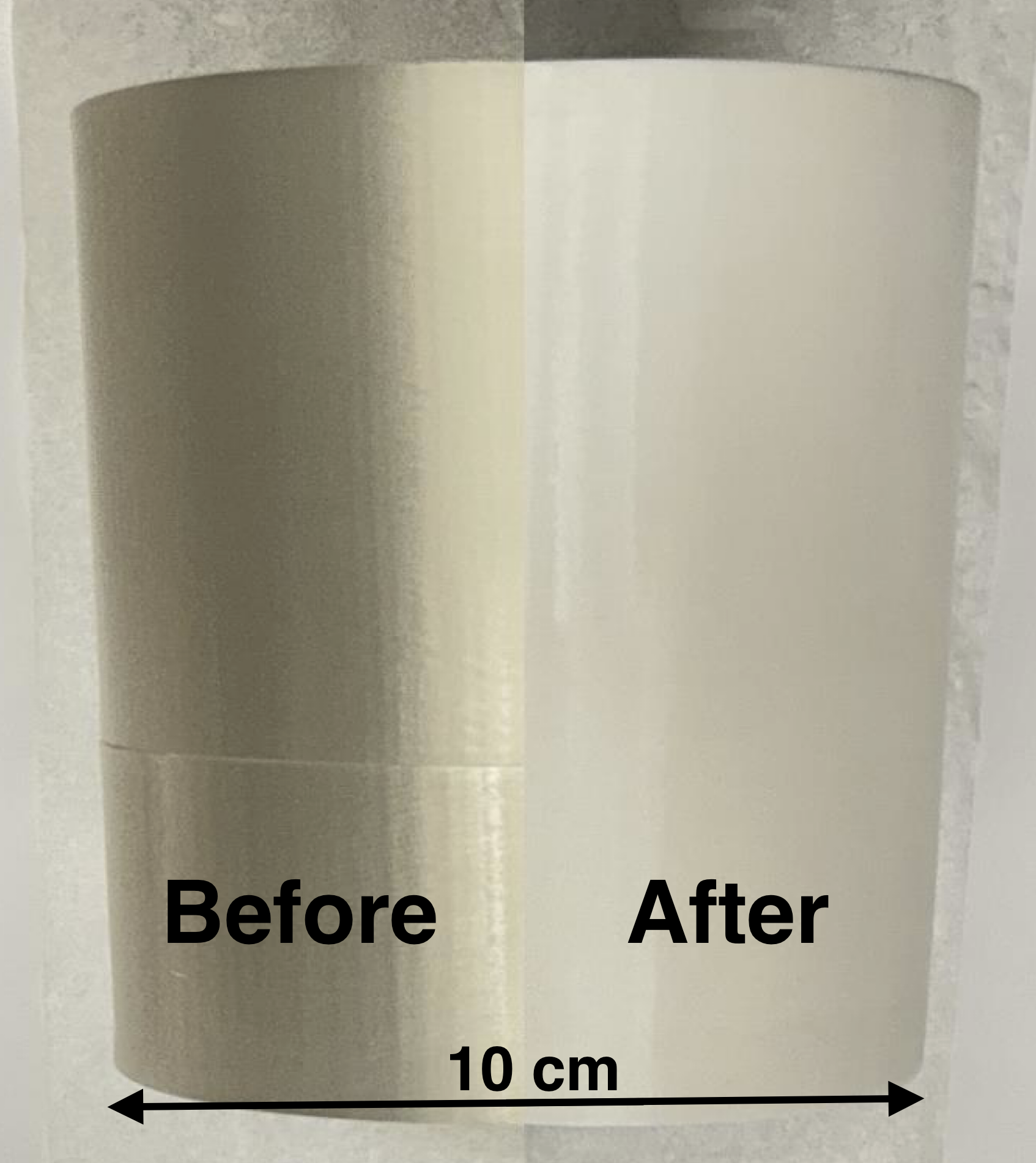}
    \end{minipage}%
    \hfill%
    \begin{minipage}[c]{0.43\textwidth}
        \vspace{0pt}
        \centering
        \includegraphics[width=\linewidth]{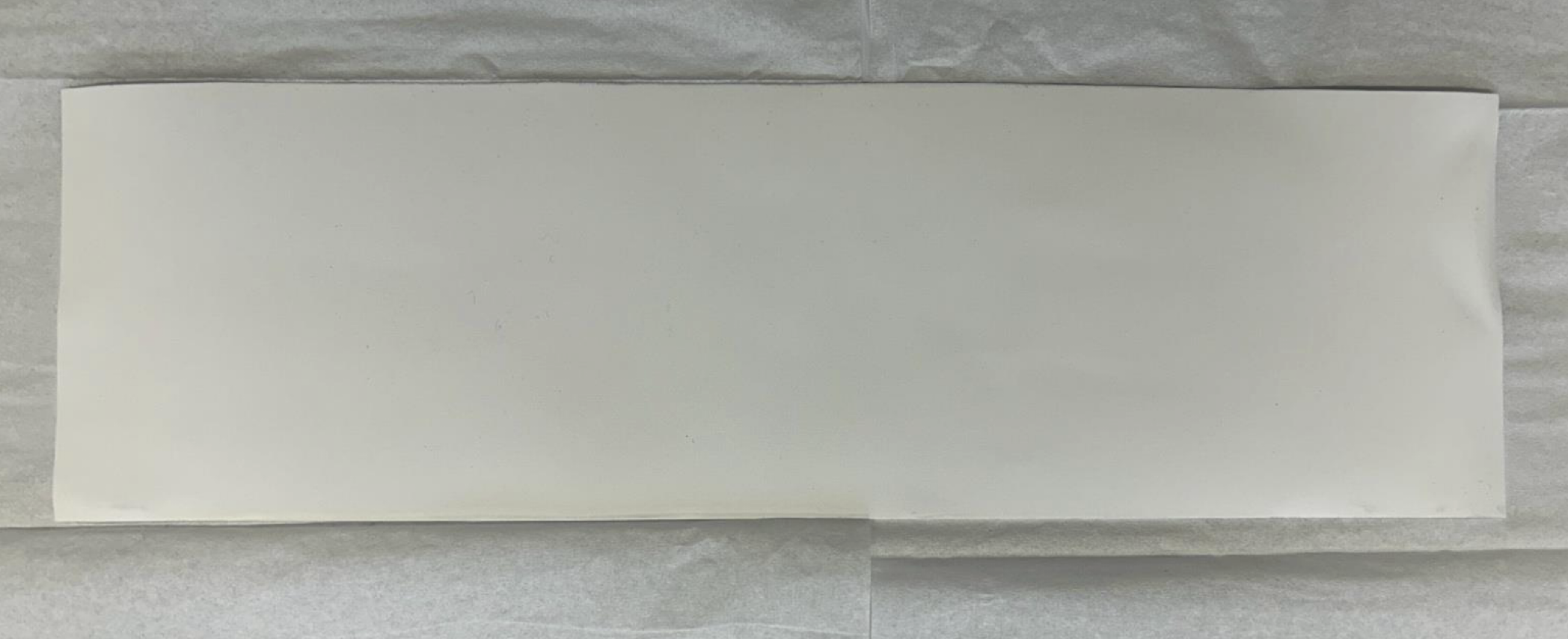}

        \vspace{0.3cm}

        \includegraphics[width=\linewidth]{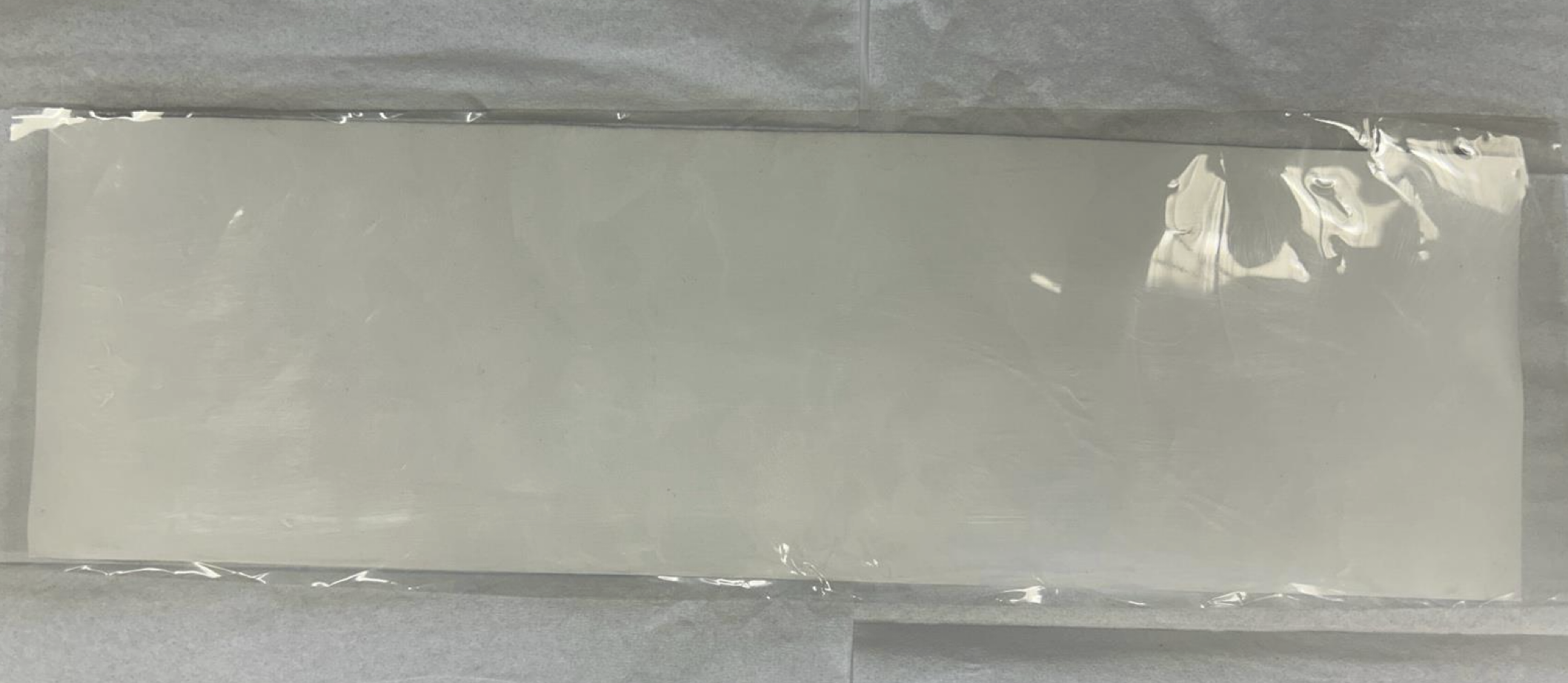}
    \end{minipage}
\caption{Picture of the 3D-filament printed PEN before and after crystallization on the left. After annealing, the PEN appears white; crystallinity was not quantified. PET sample uncoupled (upper picture) and coupled (bottom picture) with the PEN foil on the right.}
\label{fig:pictures_samples}
\end{figure}

For the VUV measurements in the liquid argon setup, see Sect.~\ref{Characterization in LAr}, the PET sample was also tested mechanically coupled to a $75~\mathrm{\mu m}$-thick PEN Q51 foil from Goodfellow Advanced Materials \cite{PEN_foil}. The PEN foil was sanded with sandpaper grit $1600$ in random directions. Fig.~\ref{fig:pictures_samples} shows the 3D-filament printed PEN before and after the annealing process on the left, and the PET sample coupled and not coupled with the PEN foil on the right.

In this work a sample of $250~\mathrm{\mu m}$-thick TTX from Donaldson \cite{Tetratex} was also characterized in LAr illuminated with blue light to serve as a reference for comparison with the novel materials.

\section{Characterization in LAr}
\label{Characterization in LAr}

The samples were characterized in LArS, a dedicated LAr setup at the University of Zurich \cite{GabrielaArticle,Baudis2015Enhancement}. Their responses to $450$\,nm LED light and $128$\,nm LAr scintillation were compared with two reference absorbers, see Section~\ref{description_measurements}. The observables are the mean number of photoelectrons (PE) per triggered LED pulse and the PE yield associated with the LAr scintillation light generated by the decays of an $\alpha$ source, respectively. The blue-light measurements were also used to infer reflectivity in LAr through optical simulations, as we detail in Section~\ref{Monte_Carlo_simulation}.

\subsection{Setup}
\label{sec:setup}

The LArS facility consists of a cryostat, $28~\mathrm{cm}$ in diameter and $50~\mathrm{cm}$ in height, coupled to an external gas handling system. Grade-6.0 argon (nominal purity \SI{99.9999}{\percent}) was liquefied and kept cold inside the cryostat by means of a coiled copper tube cooled by flowing liquid nitrogen. The cooling and system stabilization were controlled by a specifically developed programmable logic controller (PLC), described in Appendix \ref{appendix_PLC}. 

The setup is equipped with a 3-inch Hamamatsu R11065-10 photomultiplier tube (PMT) featuring a \ce{MgF2} window, which transmits both visible and VUV light down to approximately $115~\mathrm{nm}$.
For this study, a cylindrical sample holder made of black-anodized aluminum was specifically designed to accommodate the investigated reflector materials along its inner wall. The holder consists of three detachable components: a cylindrical body, a bottom endcap, and a top endcap with a circular opening of diameter $76~\mathrm{mm}$. This design allows the WLSR samples to be easily installed and replaced.
A \ce{^{241}Am} source with an activity of $30~\mathrm{Bq}$, deposited on a metallic disc of $7~\mathrm{mm}$ in diameter, is mounted on the bottom endcap. The source predominantly emits $\alpha$ particles with an energy of $5.486~\mathrm{MeV}$, producing localized scintillation in the surrounding LAr. The resulting scintillation pulses exhibit a time structure distinct from that of background events, including $\beta$ radiation from \ce{^{39}Ar} decays and $\gamma$ radiation from radioactivity in the materials used in the experimental setup and from environmental sources. A cryogenically compatible blue LED, with a peak emission wavelength of approximately $450~\mathrm{nm}$, is positioned near the source and driven by an external 50 MHz Aim-TTi TG5011A pulse generator.

\begin{figure}[htbp]
    \centering

    \begin{minipage}{0.50\textwidth}
        \centering
        \includegraphics[width=\textwidth]{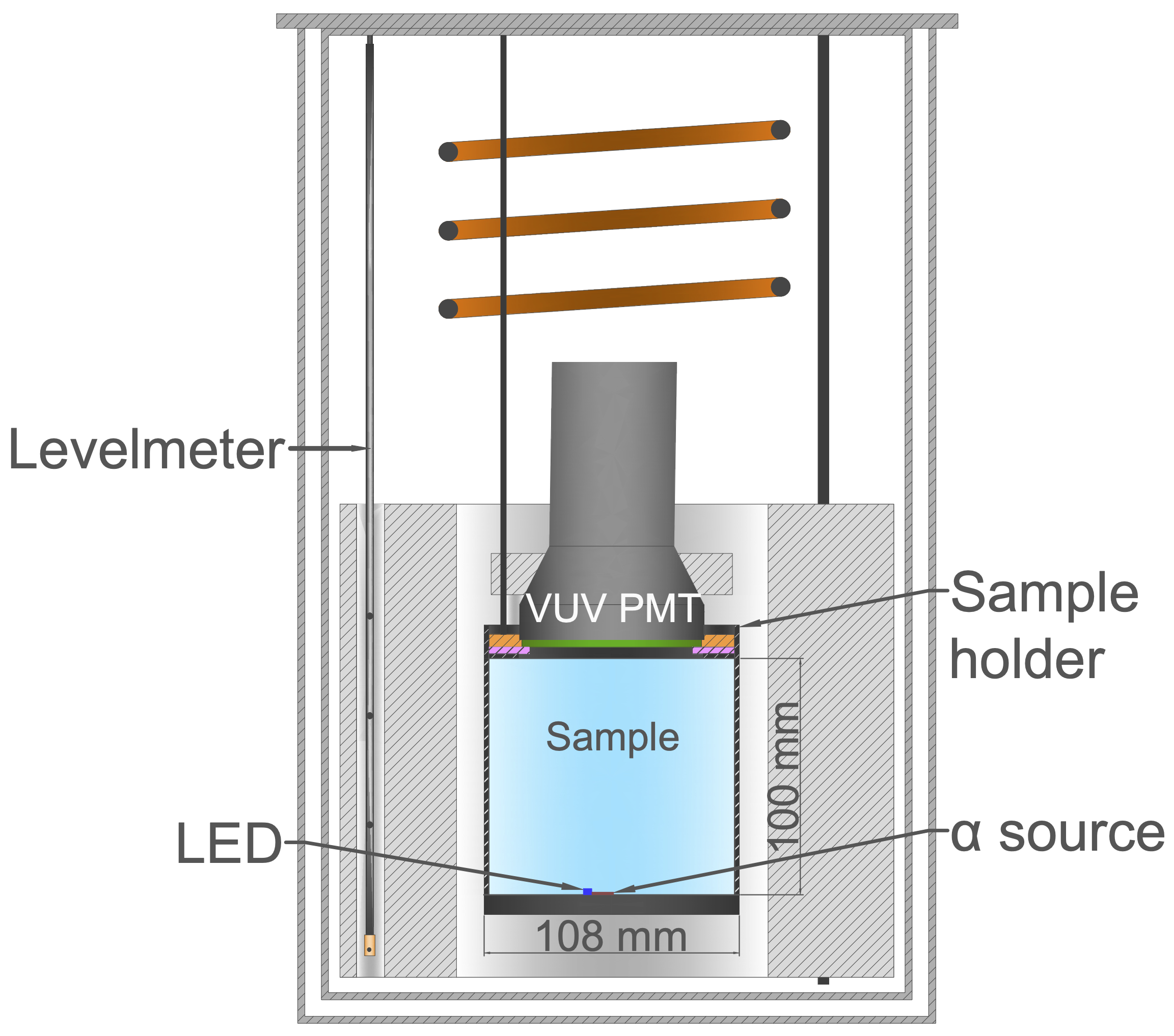}

    \end{minipage}
    \hfill
    \begin{minipage}{0.38\textwidth}
        \centering
        \includegraphics[width=\textwidth]{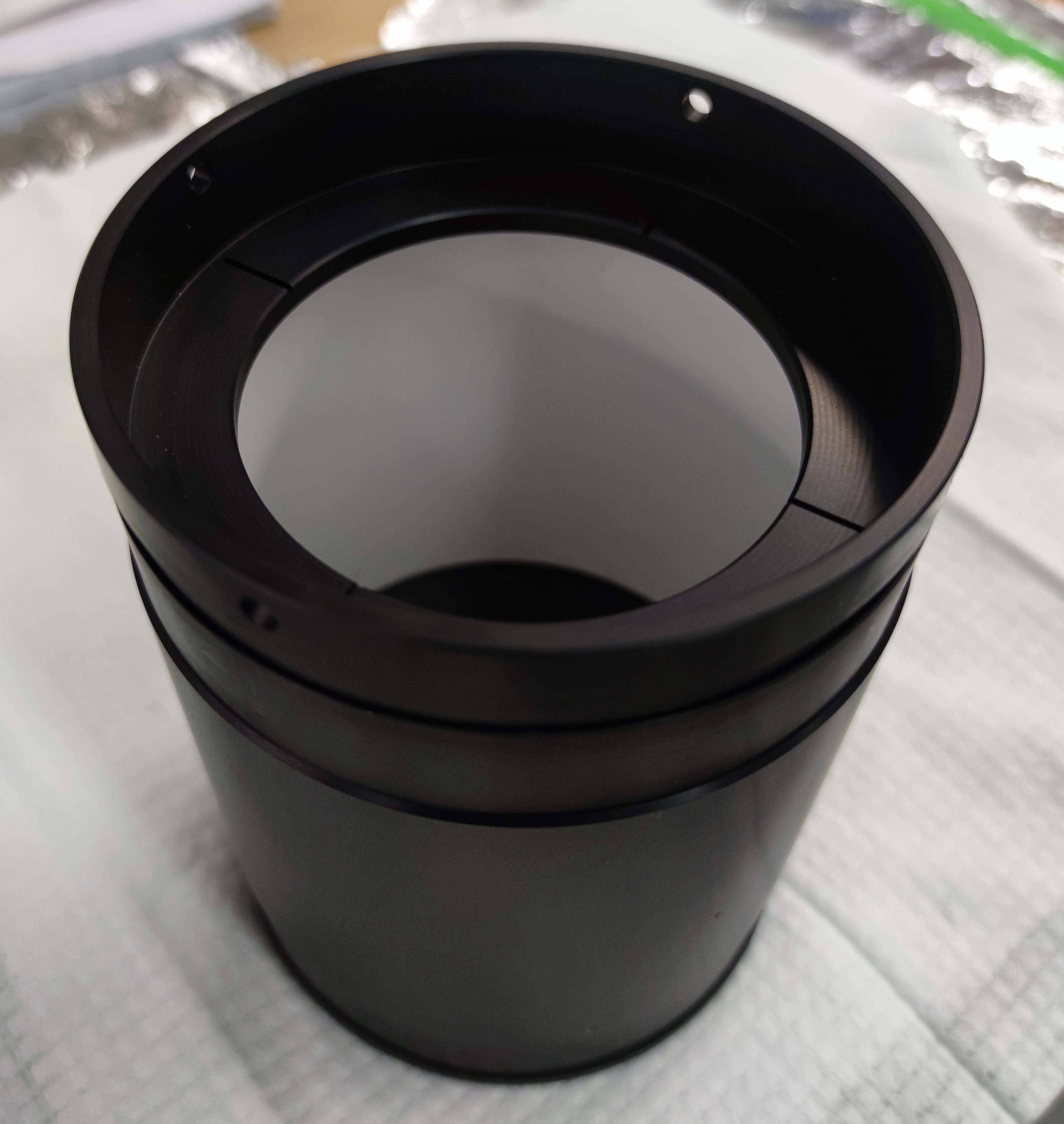}
       
    \end{minipage}
\caption{Drawing of the LArS setup on the left and picture of the black-anodized aluminum sample holder with the PET sample inside on the right.}
\label{fig:LArS}
\end{figure}

A hollow PTFE filler is placed between the cryostat's inner vessel and the active volume to reduce the amount of LAr required. The setup is equipped with a capacitive level meter to monitor the LAr level (see Appendix \ref{LArS levelmeter}), a pressure sensor, and two four-wire PT100 temperature sensors. The latter are located near the copper coil and below the PTFE filler to measure the temperatures in the gas and liquid phases, respectively.
Figure~\ref{fig:LArS} shows a schematic of the LArS setup and a photograph of the sample holder.

The PMT waveforms were acquired using a CAEN V1724 digitizer operating at a sampling rate of $100~\mathrm{MHz}$. Each waveform comprised 1500 samples, corresponding to an acquisition window of $15~\mathrm{\mu s}$, of which the first 100 samples ($1~\mathrm{\mu s}$) preceded the trigger and were used to determine the baseline. No signal amplification was applied, and the PMT was operated at a bias voltage of $1.5~\mathrm{kV}$ throughout the measurements. All waveforms were saved for subsequent processing.

\subsection{Description of the measurements}
\label{description_measurements}

Before the measurements, the PET and TTX samples were cut in tiles of $11~\mathrm{cm}\times34~\mathrm{cm}$, then wiped with isopropanol and sonicated for $\sim15~\mathrm{min}$ in a bag filled with isopropanol. The crystallized PEN, already received in the correct dimensions to fit the LArS sample holder, was wiped with a solution of deionized water and isopropanol.

To maintain a stable and high level of LAr purity, a purification system equipped with gaseous Ar recirculation pump and  GasPuri RGG5P-L getter was connected to the LArS cryostat and operated during every sample measurement. This was done to prevent impurities from decreasing LAr scintillation yield \cite{Acciarri2010Nitrogen}. Moreover, to guarantee reproducible conditions, the cryostat was evacuated to a pressure of $\sim 10^{-4}~\mathrm{mbar}$ before every new gaseous Ar filling, and LAr was evaporated at the end of each measurement. During each measurement, the LAr level was kept $5~\mathrm{cm}$ above the PMT window, to ensure the entire sample holder and PMT photocathode were always immersed in LAr. Operating LAr temperature and absolute pressure were set to $92.6~\mathrm{K}$ (measured by the PT100 located below the PTFE filler) and $1500~\mathrm{mbar}$, respectively. The PTFE filler was baked at $110~^\circ\mathrm{C}$ for $\sim2~\mathrm{h}$ and inserted into the cryostat immediately before evacuating the setup, allowing the outgassing products released by the polymer to be pumped away.

Two acquisition modes were used: an external trigger from the LED pulser for blue-light measurements, and channel self-triggering with a predefined negative-edge threshold, set to $30~\mathrm{ADC~counts}$ from the baseline, for VUV measurements. For both modes, the integration window chosen was the entire acquisition window ($15~\mathrm{\mu s}$). The LED wavelength is close to the emission peaks of TPB and PEN, around $425~\mathrm{nm}$ \cite{GabrielaArticle}, and probes the visible-light response in this spectral region. For the VUV measurements, PET was tested both alone and mechanically coupled to the PEN foil. Crystallized PEN was tested without an additional wavelength-shifting layer.

The detected PE yield was compared with that measured using black metal velvet (BMV) from Acktar as a reference absorber. The bare black-anodized aluminum (BAA) holder was also measured to assess its response and its potential contribution when reflector samples were installed. Samples covered the cylindrical inner wall, while the bottom and top endcaps remained uncoated.

Waveforms were analyzed using the dedicated LArS analysis software\footnote{The software is available at \url{https://github.com/gloriasenatore/LArS_analysis}.}.

\subsection{PMT calibration}
\label{sec:PMT calib}

\begin{figure}[htpb]
    \centering

    \begin{minipage}{0.48\textwidth}
        \centering
        \includegraphics[width=\textwidth]{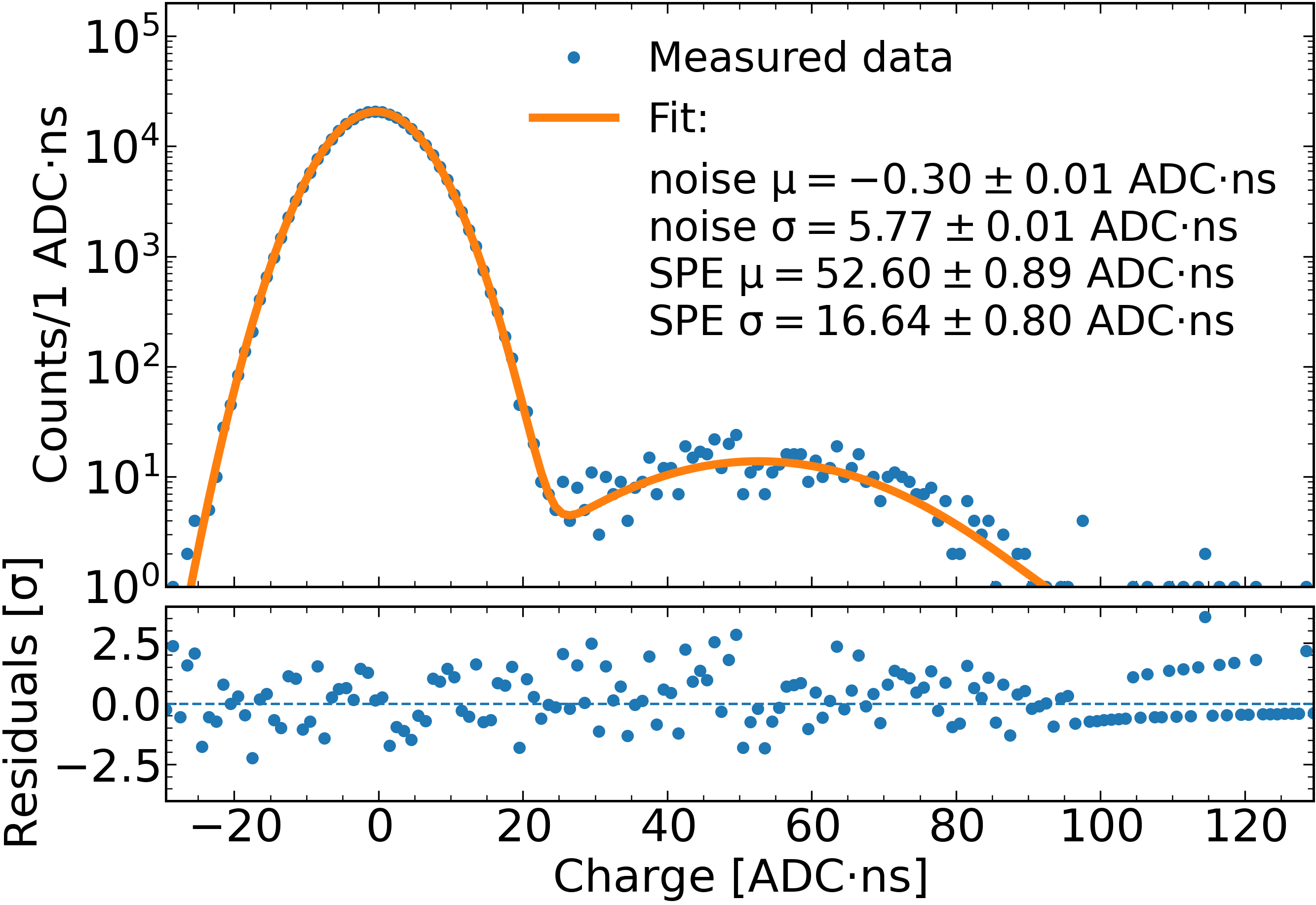}      
    \end{minipage}
    \hfill
    \begin{minipage}{0.48\textwidth}
        \centering
        \includegraphics[width=\textwidth]{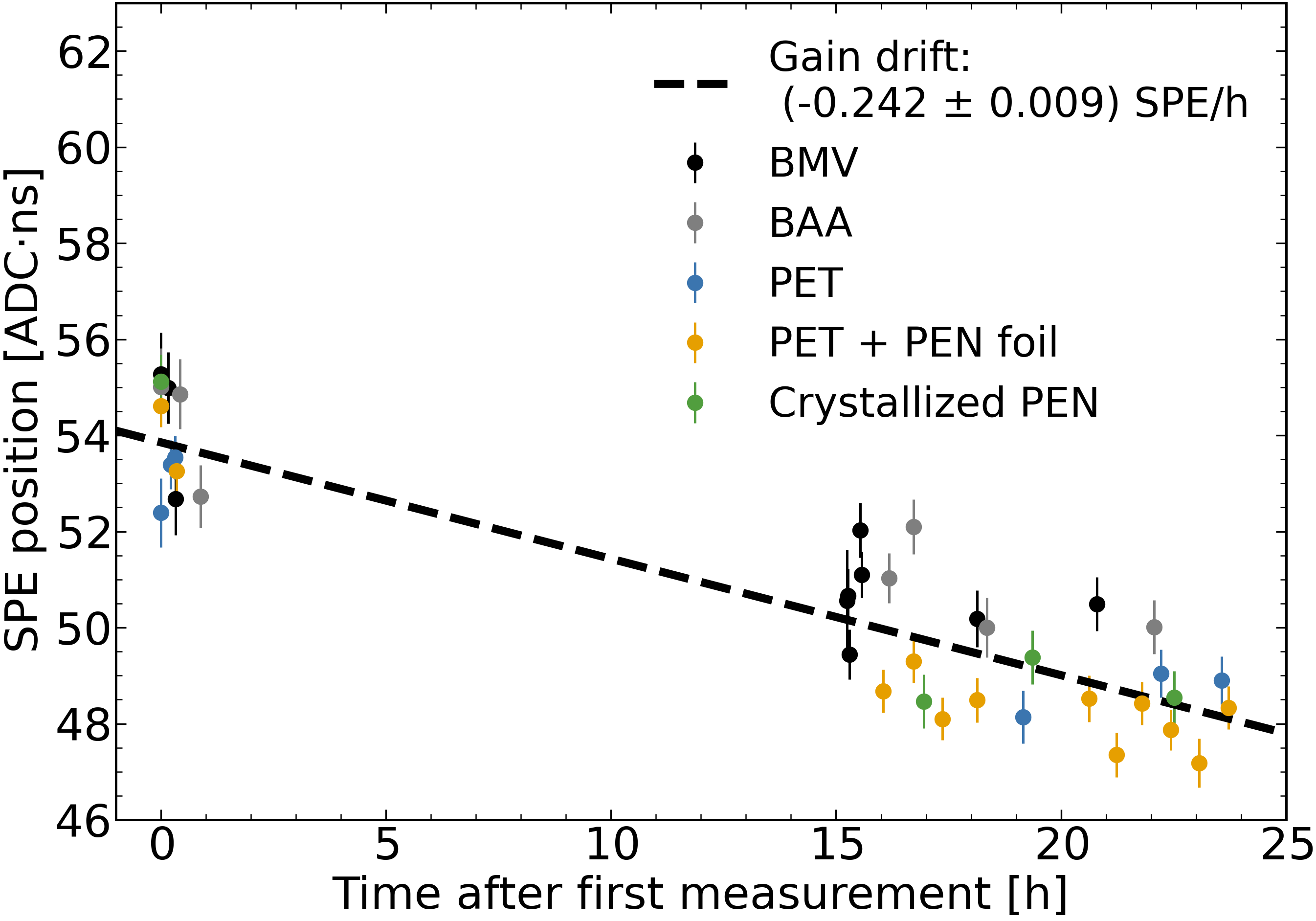}
       
    \end{minipage}
\caption{PMT calibration spectrum and fit (left), and evolution of the SPE position during the measurement campaign (right).}
\label{fig:PMT}
\end{figure}

Before each data acquisition, the PMT was calibrated using LED pulses at a repetition rate of $10~\mathrm{kHz}$ and a pulse width of $50~\mathrm{ns}$. The pulser driving the LED was set with low and high voltage levels of $0.50~\mathrm{V}$ and $1.38~\mathrm{V}$, respectively, corresponding to a voltage swing of $0.88~\mathrm{V}$. To obtain the calibration spectrum shown in the left plot of Fig.~\ref{fig:PMT}, where noise peak, single-photoelectron (SPE) peak and some events corresponding to the double-photoelectron (DPE) peak are visible, pulses were integrated in an interval of $150~\mathrm{ns}$. The SPE position was found by fitting the calibration spectrum with the function
\begin{equation}
\begin{aligned}
\mathcal{F}(&\mu_{noise}, \sigma_{noise}, \Delta, \sigma_{SPE},
 A_{noise}, A_{SPE}, A_{DPE}) \\
&= A_{noise} \cdot \mathcal{G}_{noise}(\mu_{noise}, \sigma_{noise}) \\
&\quad + A_{SPE} \cdot \mathcal{G}_{SPE}(\mu_{noise} +\Delta, \sigma_{SPE}) \\
&\quad + A_{DPE} \cdot \mathcal{G}_{DPE}(\mu_{noise}+2\cdot\Delta, \sqrt{2}\cdot\sigma_{SPE}),
\end{aligned}
\end{equation}
where $A\cdot\mathcal{G}(\mu,\sigma)$ is a Gaussian with mean $\mu$, standard deviation $\sigma$ and normalization $A$, and $\Delta$ is the distance between two adjacent peaks. To check the PMT linearity of response, a fit with distance between peaks and $\sigma_{DPE}$ let free was also performed: the SPE position was found in good agreement with the previous method.

A PMT gain drift of $-0.24~\mathrm{ADC\cdot ns}$ per hour in the SPE position was observed, see Figure~\ref{fig:PMT}, right. The PMT was calibrated before each acquisition to account for changes in gain between datasets.

\subsection{Blue-light measurements data analysis}

\begin{figure}[htpb]
    \centering

    \begin{minipage}{0.48\textwidth}
        \centering
        \includegraphics[width=\textwidth]{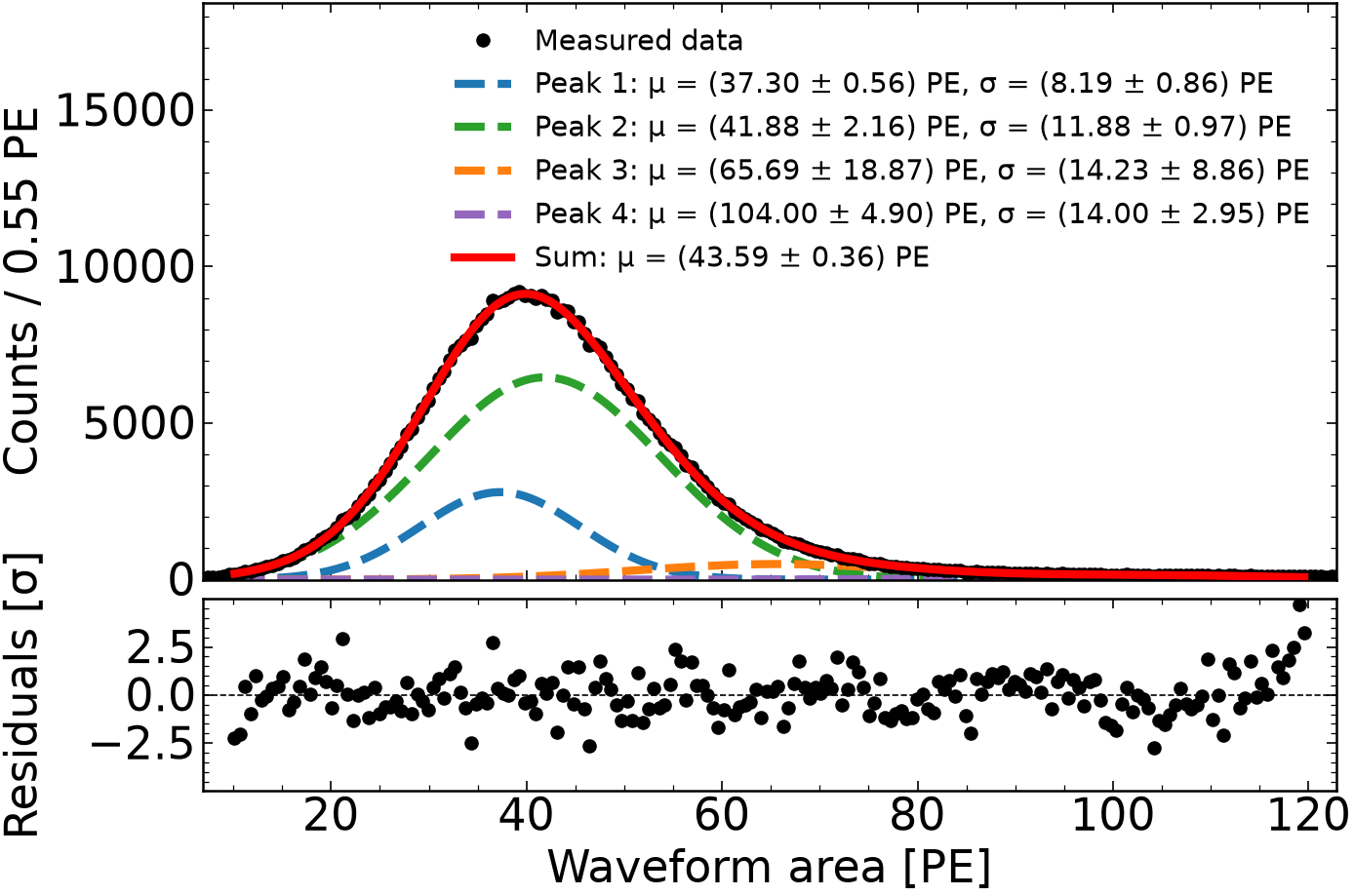}

    \end{minipage}
    \hfill
    \begin{minipage}{0.48\textwidth}
        \centering
        \includegraphics[width=\textwidth]{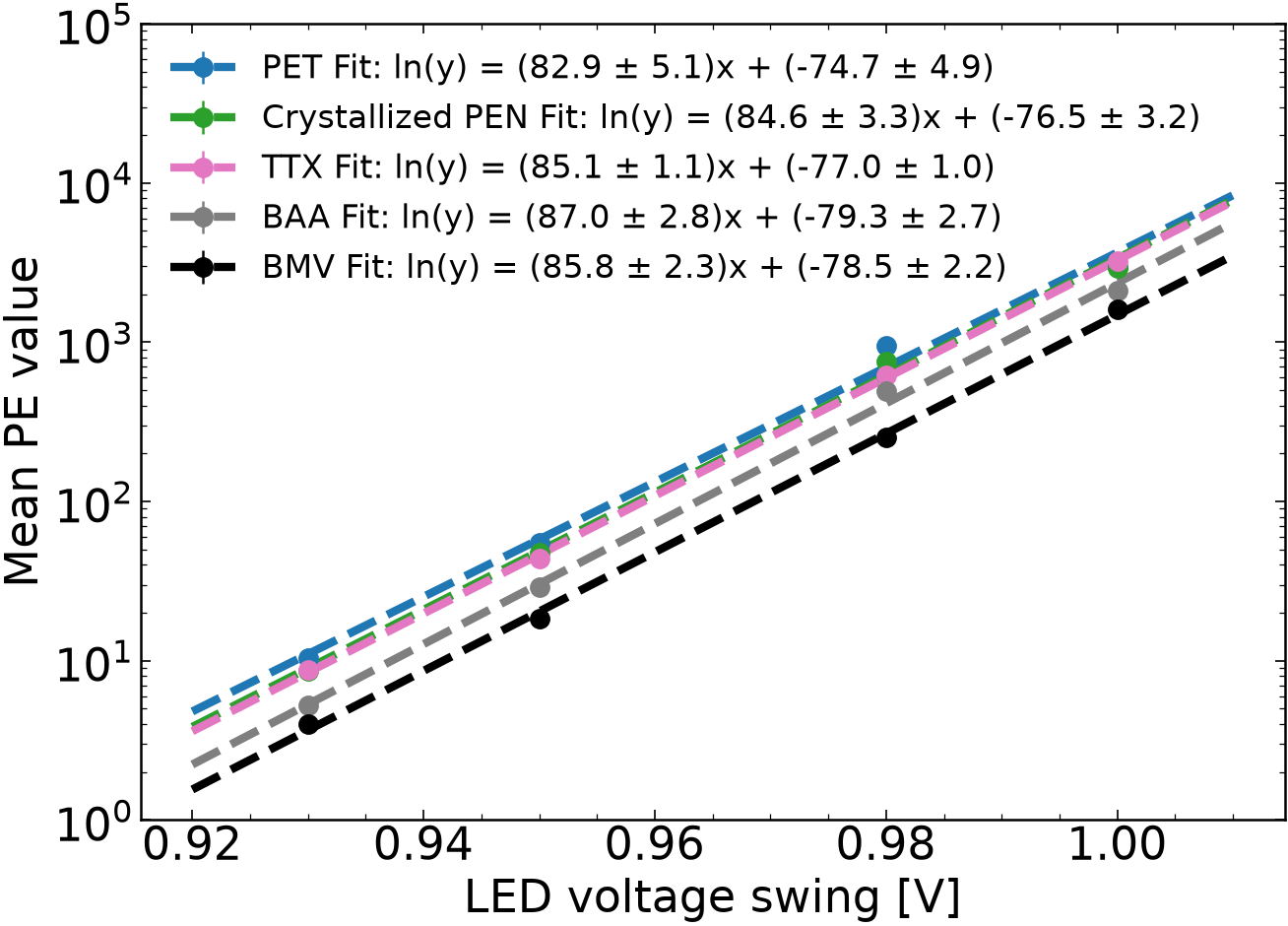}
       
    \end{minipage}
\caption{Fitted blue-light measurements waveform-integral spectrum on the left. Mean PE values as a function of the LED voltage swing on the right: error bars representing the uncertainties from the fit of the waveform-integral spectra are included but are too small to be visible on the log-linear plot. For each material, the data points are fitted with an exponential function, and results from the fit are reported, where in the legend $y$ represents the mean PE value and $x$ the LED voltage swing.}
\label{fig:LED_bias_vs_mean_PE}
\end{figure}

For each sample, a scan at different LED intensities was performed following the PMT calibration described in Section \ref{sec:PMT calib}. The LED pulse repetition rate and width were set to $1~\mathrm{kHz}$ and $200~\mathrm{ns}$, respectively. The pulser low voltage level was kept constant at $0.50~\mathrm{V}$, while the high voltage level was varied across four settings: $1.43~\mathrm{V}$, $1.45~\mathrm{V}$, $1.48~\mathrm{V}$, and $1.50~\mathrm{V}$, corresponding to voltage swings between $0.93~\mathrm{V}$ and $1.00~\mathrm{V}$. At each setting, $5\cdot10^5$ events were acquired.
The acquired waveforms were integrated to obtain a waveform-integral spectrum, which was then calibrated using the PMT single-photoelectron (SPE) response to express the integrals in photoelectrons (PE). The left panel of Fig.~\ref{fig:LED_bias_vs_mean_PE} shows an example of this spectrum for a pulser high-level voltage of $1.45~\mathrm{V}$.

At pulser high-level voltages of $1.43~\mathrm{V}$ and above, a single Gaussian did not adequately describe the waveform-integral spectra. Additional structure was observed, and individual waveforms showed a secondary feature preceding the primary LED pulse. Oscilloscope measurements also revealed distortions of the LED drive signal relative to a rectangular pulse, which may explain the multiple-pulse structure.

The observed spectra were therefore modeled as a Gaussian mixture. The estimation of the mean PE value with a given sample was obtained from this multi-Gaussian model as the area-weighted mean across all fitted components: 
\begin{equation}
\bar\mu = \frac{\sum_i a_i \sigma_i \mu_i}{\sum_i a_i \sigma_i}
\end{equation}
where, for the $i$-th component, $a_i$ is the fitted amplitude, $\sigma_i$ the standard deviation, and $\mu_i$ the mean. Each component is weighted by its area $a_i\sigma_i$ rather than its amplitude, so that a narrow, high-amplitude component does not outweigh a broader one that could contain more events. The uncertainty on $\bar\mu$ was estimated by using the full covariance matrix returned by the fit: 
\begin{equation}
  \sigma_{\bar\mu}^2 = J^T C J,  
\end{equation}
where $C$ is the covariance matrix of the fit parameters $p_k$ (the amplitudes, means and standard deviations of all $N$ components), and $J$ is the vector of partial derivatives $J_k = \partial \bar\mu / \partial p_k$, evaluated at the best fit parameter values. The uncertainty from the fit was combined in quadrature with the independent uncertainty on the SPE calibration to obtain the total uncertainty on the mean PE. The resulting uncertainties were cross-validated against a model-independent estimate of the mean and uncertainty computed directly from the binned data, and against a bootstrap resampling of raw events, both in agreement with the fit results.

Table~\ref{tab:LArS_results_blue} reports the mean PE for each tested material at a pulser voltage swing of $0.95~\mathrm{V}$. The table also reports the result obtained with TTX for comparison.

\begin{table}[htbp]
    \centering
    \caption{Blue-light PE yields at an LED pulser voltage swing of $0.95~\mathrm{V}$. The last column gives the yield relative to BMV. TTX is included as a reference reflector.}
    \label{tab:LArS_results_blue}
    \smallskip

    \begin{tabular}{c|c|c|c}
        \hline
        \textbf{Sample} & \textbf{Type} &
\textbf{Mean PE value} & \textbf{Ratio} \\
        \hline
        BMV & absorber & $18.3\pm0.3$ & $1$ \\

        BAA & absorber & $29.0\pm0.4$ & $1.58\pm0.04$ \\
        
        PET & reflector & $55.1\pm0.8$  & $3.01\pm0.07$ \\
     
        Crystallized PEN & reflector & $47.4\pm0.9$ & $2.59\pm0.06$ \\
        
        TTX & reflector & $43.6\pm0.4$  & $2.38\pm0.05$\\
        \hline
        
    \end{tabular}
\end{table}

Figure~\ref{fig:LED_bias_vs_mean_PE} (right) shows the mean PE yield versus LED voltage swing, with an exponential fit for each configuration. The ordering of the materials in Table~\ref{tab:LArS_results_blue} is preserved across the tested settings. Fit uncertainties are included but are too small to resolve on the logarithmic scale.

\subsection{VUV measurements data analysis}
\label{VUV_measurements}

\begin{figure}[htbp]
    \centering
    \includegraphics[width=0.7\textwidth]{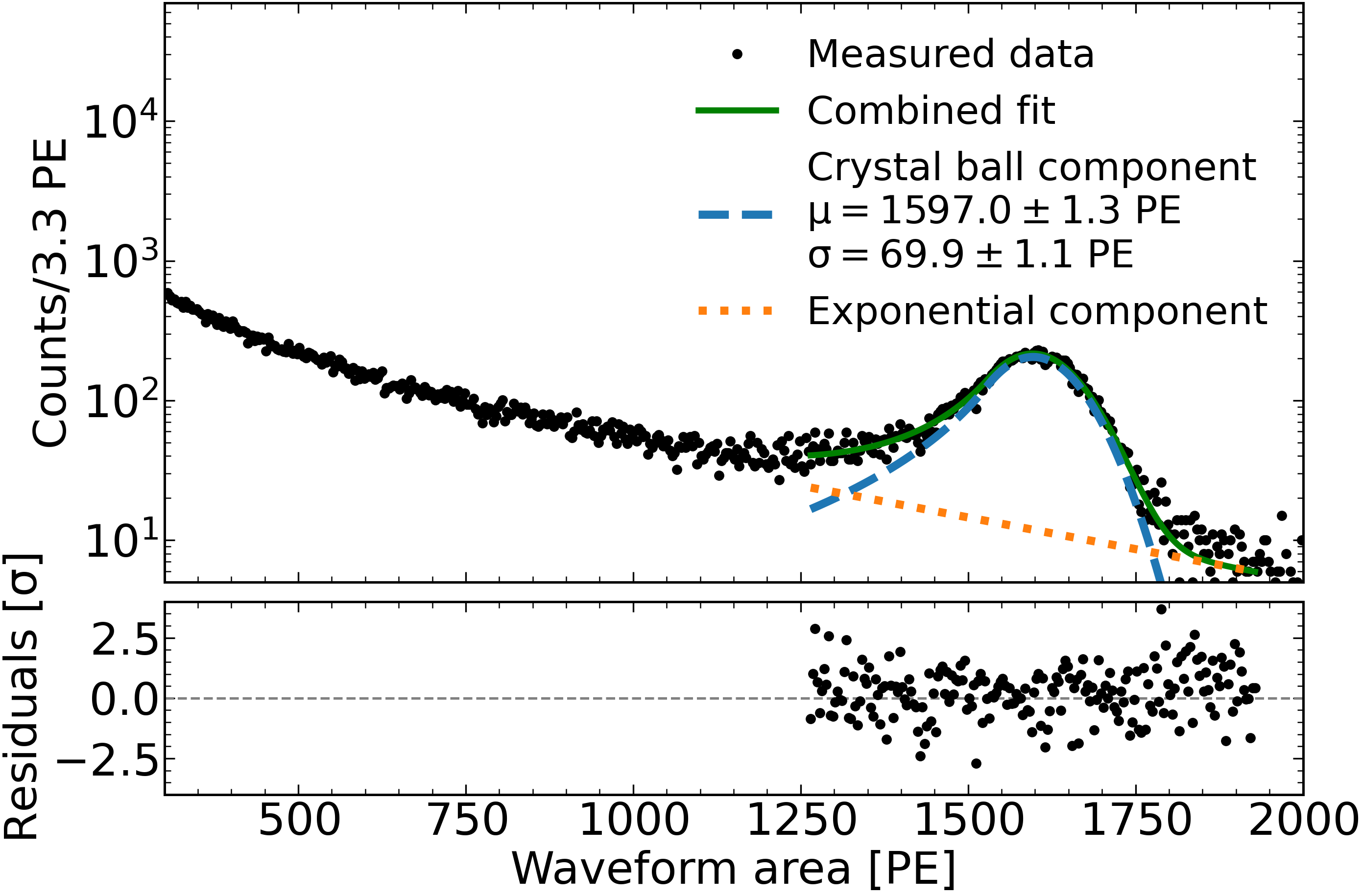}
    \caption{Example VUV waveform-integral spectrum with a Crystal Ball signal model and exponential background.}
    \label{fig:alpha_bump_plot}
\end{figure}

For the VUV measurements, batches of $10^6$ waveforms were acquired regularly over two days, with a PMT calibration before each acquisition. Waveform integrals were converted to PE using the SPE calibration. The $\alpha$-peak signal includes scintillation photons reaching the PMT directly, reflected photons, and, where applicable, wavelength-shifted photons. Each spectrum was fitted with a Crystal Ball function for the signal and an exponential background to estimate the PE yield. The Crystal Ball core centroid corresponds to the PE yield obtained.
Figure~\ref{fig:alpha_bump_plot} shows an example of a spectrum fit. 

\begin{figure}[htbp]
    \centering

    \begin{minipage}{0.48\textwidth}
        \centering
        \includegraphics[width=\textwidth]{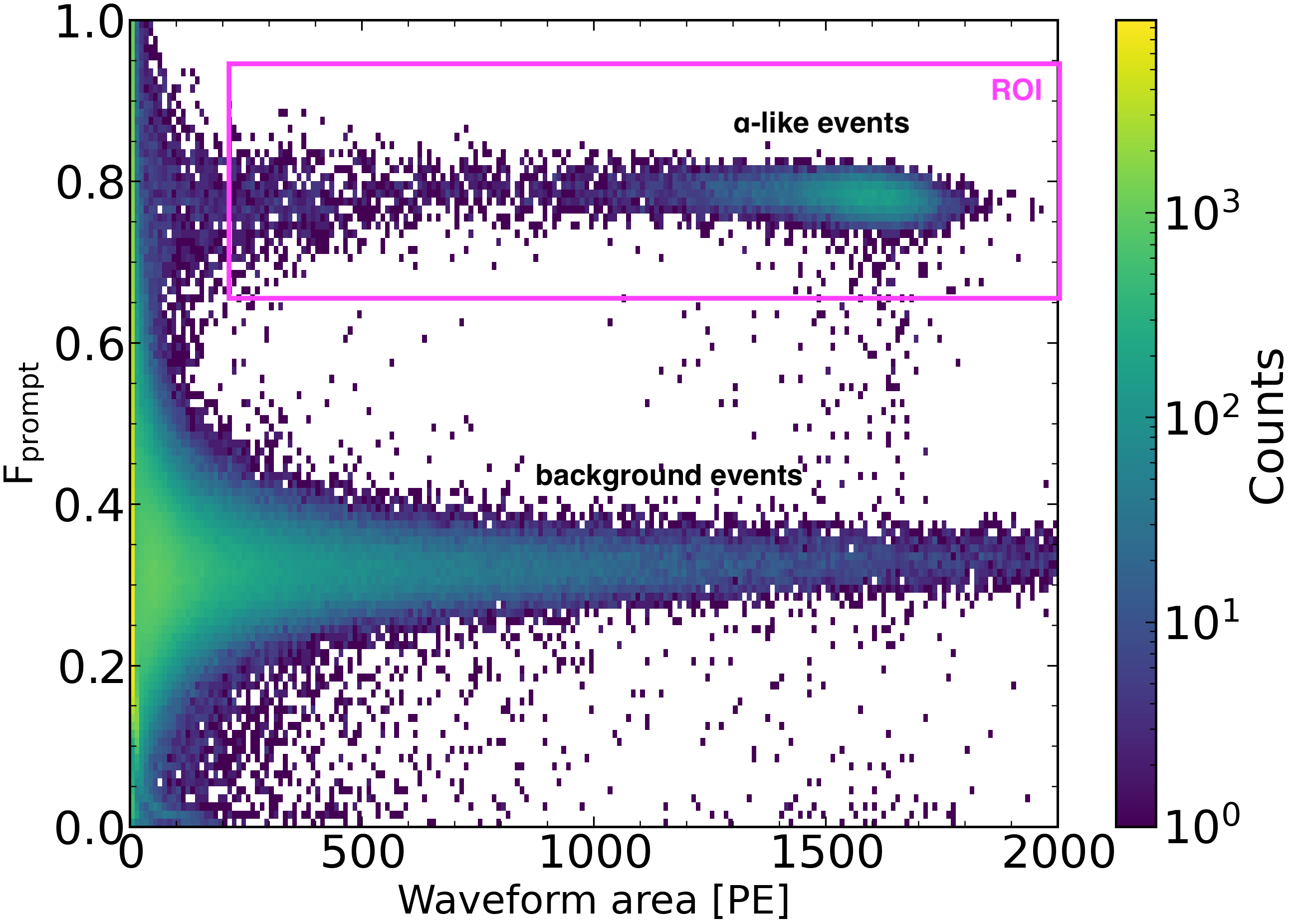}

    \end{minipage}
    \hfill
    \begin{minipage}{0.48\textwidth}
        \centering
        \includegraphics[width=\textwidth]{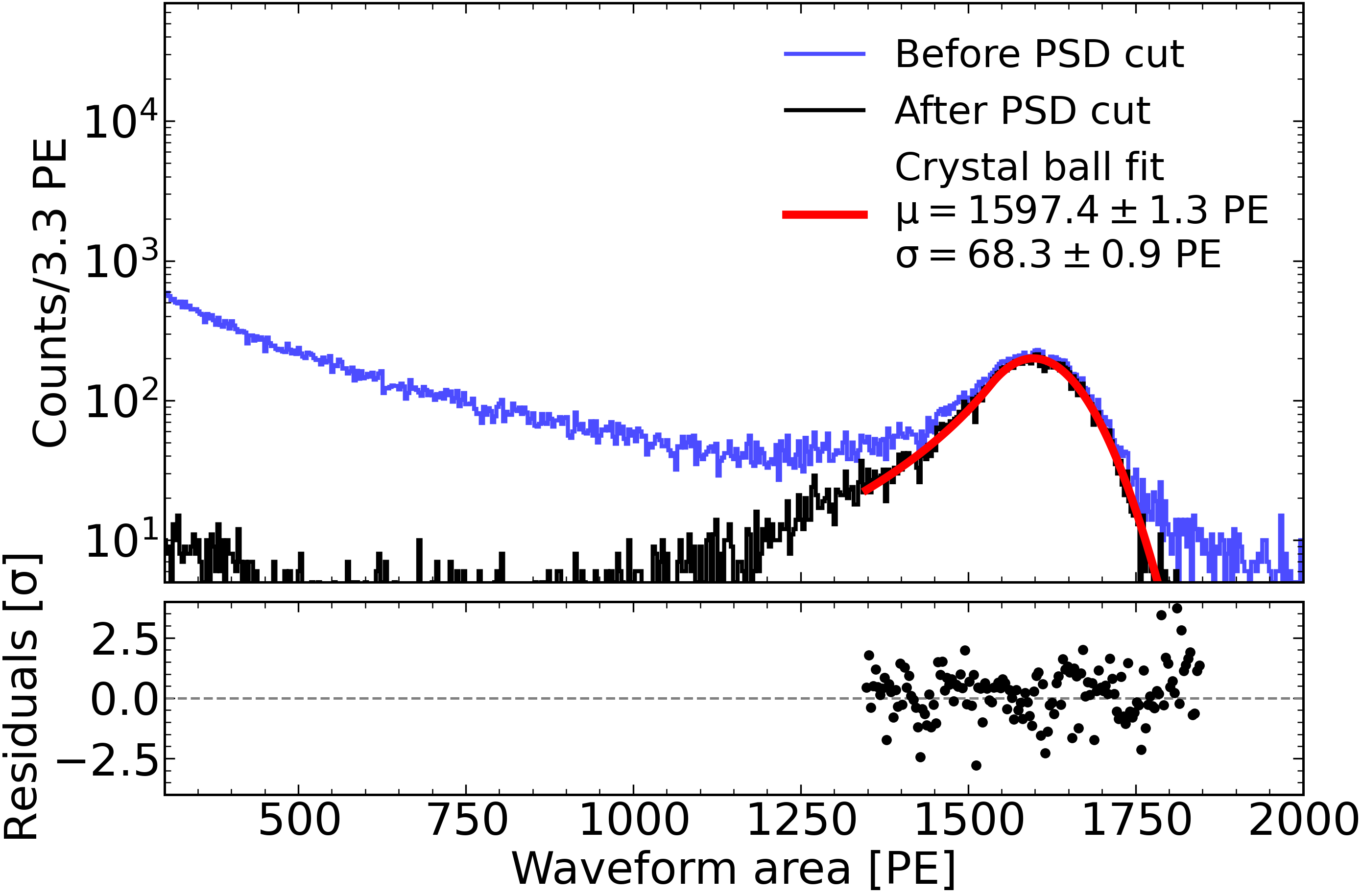}
       
    \end{minipage}
\caption{Two-dimensional histogram of the fraction of prompt light, $F_{prompt}$, and integrated waveforms value on the left: the region of interest (ROI) used to select $\alpha$-like events is shown. Fitted waveform-integral spectrum performed on $\alpha$-like events only on the right.}
\label{fig:2D_PSD_plot}
\end{figure}

To cross-check the results, another analysis of the waveform-integral spectra, based on a pulse shape discrimination (PSD) method, was performed, following the approach of Ref. \cite{GabrielaArticle}. The PSD figure of merit employed is the fraction of prompt light, $F_{prompt}$, which is defined as the ratio of the charge integrated over the first $200~\mathrm{ns}$ following the scintillation-pulse onset to the total integrated charge \cite{segreto2021properties}. Since $\alpha$-like events are characterized by a fraction of prompt light of $\sim 0.8$, in contrast with the one of background events of $\sim 0.3$, this PSD cut can effectively select $\alpha$-like events, as shown in Figure~\ref{fig:2D_PSD_plot}. 

Mean PE values before and after the PSD cut were found in good agreement, and only data before the PSD cut are shown in Table~\ref{tab:LArS_results_VUV}. The reported mean PE values and uncertainties are evaluated by averaging the results from measurements acquired over the two days for each sample; the uncertainties were then combined with uncertainties on the SPE position.

\begin{table}[htpb]
    \centering
    \caption{PE yields under LAr scintillation illumination, ratios relative to BMV, and measured triplet lifetimes $\tau_t$. PET was tested alone and mechanically coupled to a PEN foil.} 
    \label{tab:LArS_results_VUV}
    \smallskip

    \begin{tabular}{c|c|c|c|c}
        \hline
        \textbf{Sample} & \textbf{Type} & \textbf{Mean PE value} &
\textbf{Ratio} & \textbf{$\tau_t$ [ns] }\\
        \hline
        BMV & absorber & $672 \pm 31$ & $1$ & $1325\pm15$ \\
       
        BAA & absorber & $669 \pm 22$ & $1.00 \pm 0.06$ & $1318\pm16$ \\

        PET & reflector & $807 \pm 58$ & $1.20 \pm 0.10$ & $1398\pm12$ \\
        
        PET + PEN foil & reflector + WLS & $1588 \pm 16$ & $2.36 \pm 0.11$ & $1414\pm3$ \\

        Crystallized PEN & WLSR & $1179 \pm 34$ & $1.75 \pm 0.10$ & $1323\pm11$ \\
        \hline
        
    \end{tabular}
\end{table}

\begin{figure}[htbp]
    \centering
    \includegraphics[width=0.7\textwidth]{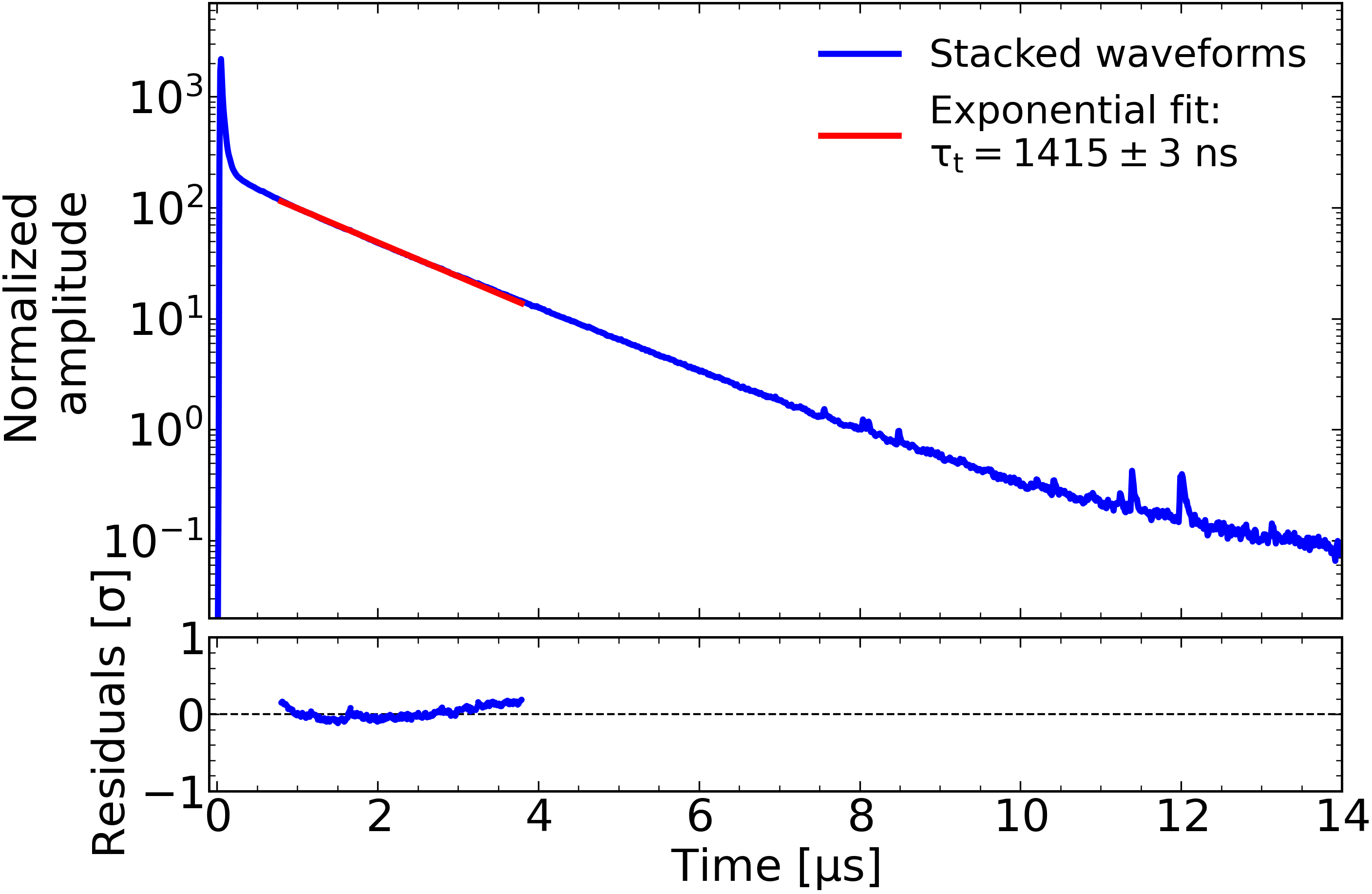}
    \caption{Averaged LAr scintillation waveform from the PET+PEN foil measurement, obtained by aligning individual waveforms to their pulse onset. The initial $1~\mathrm{\mu s}$ pretrigger baseline is omitted, and the time axis is defined relative to the scintillation onset. The exponential fit is used to determine the LAr triplet lifetime.}
    \label{fig:stacked_waveforms}
\end{figure}

To monitor LAr purity, the triplet lifetime, $\tau_t$, was extracted from averaged scintillation waveforms. Individual waveforms were baseline-subtracted and aligned to their scintillation-pulse onset before averaging. The resulting time profile was fitted with an exponential function over the interval $800$--$3800~\mathrm{ns}$, as shown in Figure~\ref{fig:stacked_waveforms}. Average values of triplet lifetime across acquisitions for each sample are reported in Table~\ref{tab:LArS_results_VUV}. The uncertainties are obtained from a weighted average of individual measurements taken at different times, with the uncertainty of each measurement derived from the corresponding fit. The measured triplet lifetimes exceed $1.30~\mathrm{\mu s}$ and are consistent with values reported by other experiments employing purified liquid argon \cite{Acciarri2010Nitrogen, Vogl2024}. The longer lifetimes in the PET and PET+PEN foil measurements may be associated with a rerouting of the Ar flow, performed before the two measurements, which forced all the injected gaseous Ar to pass through the getter before entering the LArS cryostat.

\section{Characterization in air}
\label{Characterization in air}

The room-temperature reflectance of PET and of a separately prepared crystallized PEN comparison sample was measured in air over the wavelength range $370$--$700~\mathrm{nm}$ using a PerkinElmer LAMBDA 850 UV/vis spectrophotometer equipped with a $150~\mathrm{mm}$ integrating sphere available at the Technical University of Munich. The samples were illuminated with monochromatic light at an incidence angle of $8^\circ$, and the reflected light was collected by a Hamamatsu R955 PMT.

\subsection{Procedures}

\begin{figure}[htbp] 
    \centering

    \begin{minipage}{0.48\textwidth}
        \centering
        \includegraphics[width=\textwidth]{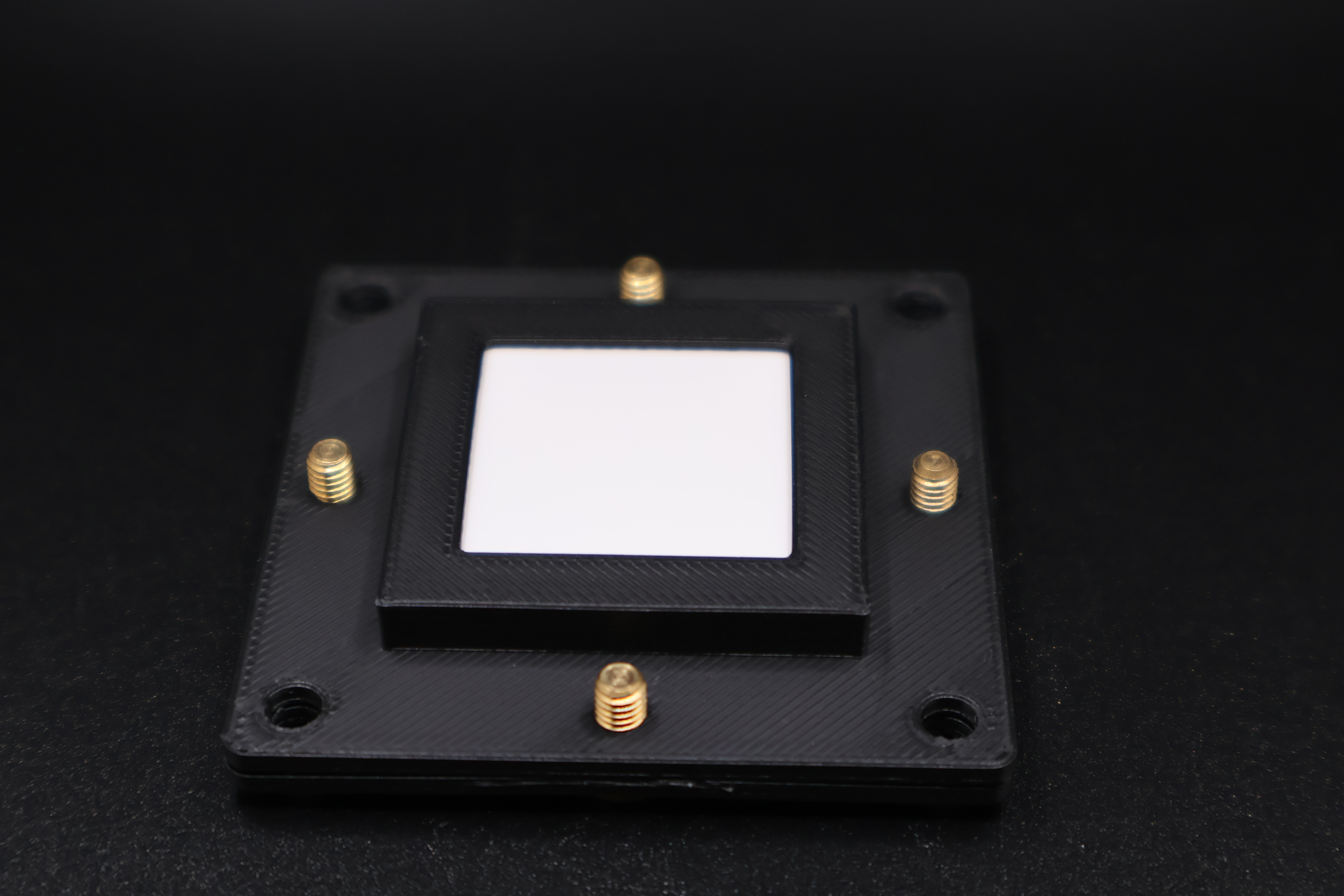}
    \end{minipage}
    \hfill
    \begin{minipage}{0.48\textwidth}
        \centering
        \includegraphics[width=0.8\textwidth, angle=270]{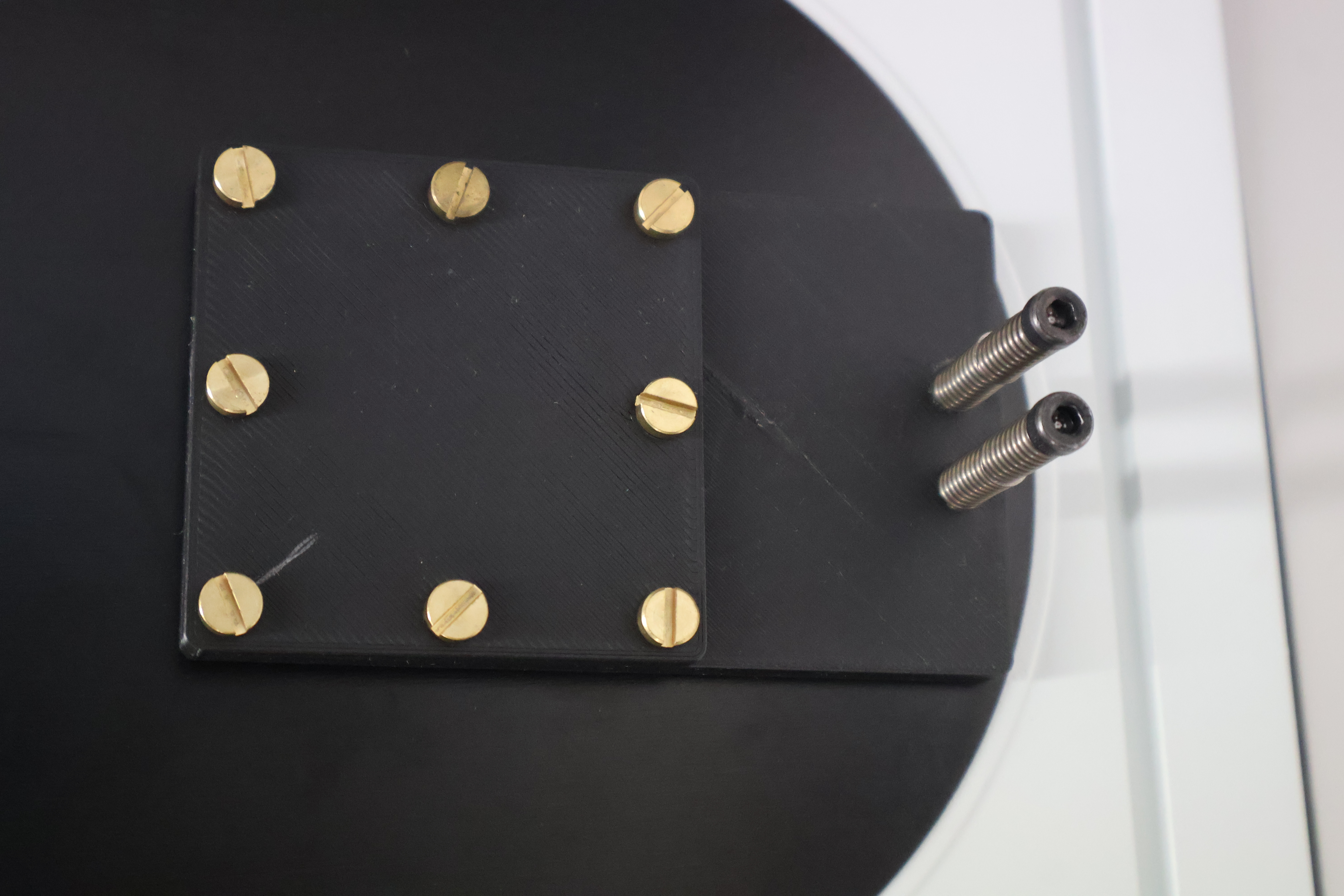}
    \end{minipage}
    
    \caption{A tile of the PET sample inside the frame designed for spectrophotometer measurements on the left. The frame is then mounted on the spectrophotometer as shown on the right.}
\label{fig:reflector_air_measurement}
\end{figure}

The samples were cut into $30\times30~\mathrm{mm}^2$ tiles, cleaned with isopropanol, and mounted in a custom 3D-printed, non-reflective holder, see Figure~\ref{fig:reflector_air_measurement}.

The spectrophotometer was calibrated using a PTFE reflectance standard. Measurements were performed in $1~\mathrm{nm}$ wavelength steps and repeated three times to assess repeatability. The resulting statistical uncertainties were negligible. Both sides of the PET sample were characterized to investigate possible differences in reflectance.

The 3D-printed crystallized PEN sample could not be measured due to its cylindrical geometry and mechanical rigidity. Instead, we measured a sample of PEN made from the same commercial granulate used for the structural components of LEGEND-200 \cite{Manzanillas_2022}. While the HPGe detector holders in LEGEND-200 were produced by injection molding, the crystallized sample investigated in this work was formed by allowing a portion of the molten material to cool in air, forming the crystalline sample. This separately prepared sample provides a comparison, although its processing differs from that of the printed cylinder. Additional measurements were performed on the PEN foil used in the LArS VUV measurements (Section~\ref{VUV_measurements}) and the black metal velvet (BMV) absorber.

Total reflectance was measured first. The integrating sphere was then reconfigured to exclude the specularly reflected light, and the instrument was recalibrated to measure the diffuse component. The specular reflectance was obtained by subtracting the diffuse reflectance from the total reflectance.

\subsection{Results}

\begin{figure}[htpb]
    \centering

    \begin{minipage}{0.48\textwidth}
        \centering
        \includegraphics[width=\textwidth]{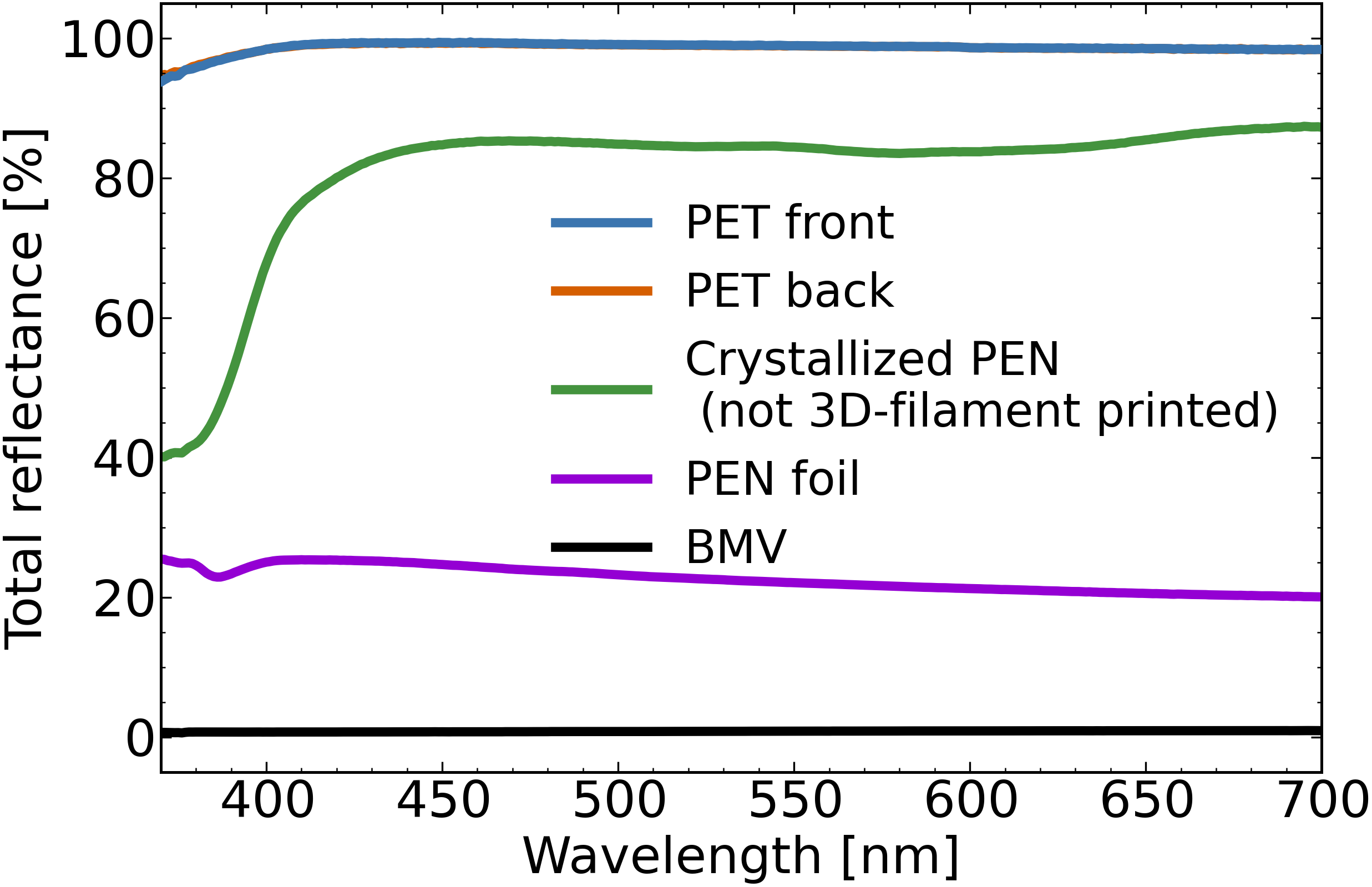}
    \end{minipage}
    \hfill
    \begin{minipage}{0.48\textwidth}
        \centering
        \includegraphics[width=\textwidth]{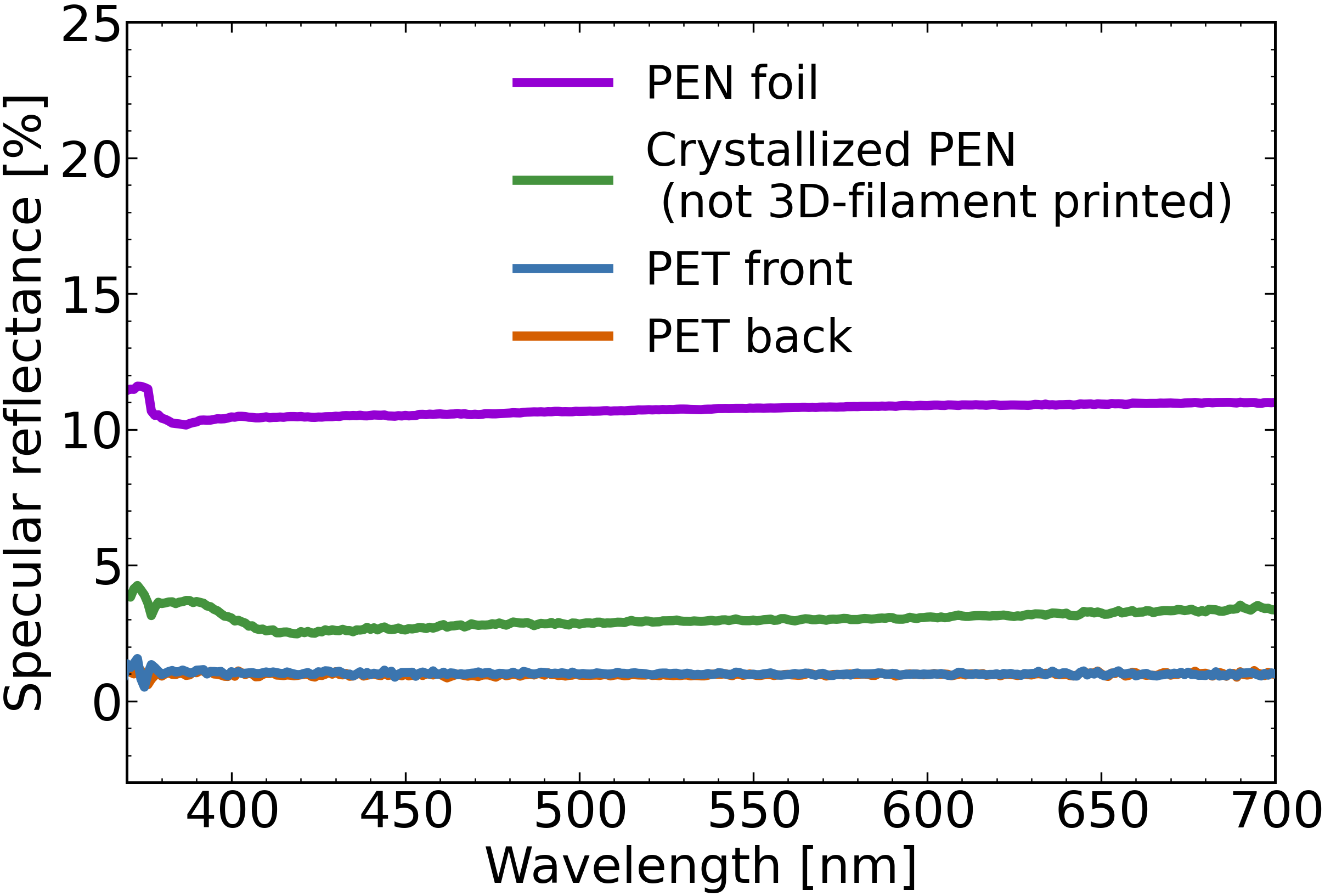}
    \end{minipage}
    
    \caption{Total (left plot) and specular (right plot) reflectance in air at room temperature of the samples tested as a function of the wavelength of the light used to illuminate the samples. The crystallized PEN (not 3D-printed) refers to the proxy sample for crystallized PEN.}
\label{fig:air_measurements}
\end{figure}

The measured total and specular reflectance spectra are shown in Figure~\ref{fig:air_measurements}. Both sides of the PET sample exhibit high total reflectance, approaching $98\%$ in the investigated wavelength range. The specular component of PET is approximately $1\%-2\%$, indicating predominantly diffuse reflection. In contrast, the PEN foil exhibits larger specular contributions. The proxy sample for crystallized PEN shows total reflectance $\gtrsim80\%$ starting from around $420~\mathrm{nm}$, and specular reflectivity around $3\%-4\%$.

The PEN foil shows lower visible reflectance than PET, consistent with previous measurements \cite{GabrielaArticle}. The measured reflectance spectra provide reference values for the optical modeling of the LArS measurements described in the following section.

\section{LArS Monte Carlo simulation}
\label{Monte_Carlo_simulation}

A Geant4-based optical simulation \cite{Agostinelli2003} of the LArS setup was developed to infer sample reflectivities at $450~\mathrm{nm}$ from the LED measurements.

The BMV measurement was used to calibrate the effective blue photon yield, and the BAA measurement constrained the holder response. Sample reflectivities were then adjusted to reproduce the measured blue-light PE yields. The resulting value for the PET sample was also used in a consistency check with the VUV measurement.

\subsection{Geometry implementation and optical parameters}

The geometry was adapted from the previous LArS simulation \cite{GabrielaArticle} to include the new sample holder dimensions. Sample thicknesses were included because they affect the optical geometry.

Regarding the LAr optical parameters, refractive indices of LAr were kept as in the previous work (Sellmeier coefficients measured in \cite{Babicz2020}). The resulting Rayleigh scattering length at $450~\mathrm{nm}$ is on the order of a few hundred meters, negligible given the setup size. The absorption length of visible light in LAr is expected to be large, although its precise value at the wavelength of interest is not well established \cite{Neumeier2012, AguilarArevalo2022}. Values from $10~\mathrm{m}$ up to $100~\mathrm{m}$ were tried in the simulation.

The PMT QE ($QE_{blue}$) was set to $27\%$ at $450~\mathrm{nm}$, according to the PMT data sheet (data provided by Hamamatsu). No uncertainty was assigned to it, considering that this parameter is degenerate with the effective blue photon yield, $N_{eff}$, as explained in \ref{abs_sim}. To estimate possible effects due to the PMT window reflectivity, the refractive indices of \ce{MgF2} at $450~\mathrm{nm}$ were investigated \cite{Dodge1984}. From these values, we estimated that the PMT window visible reflectivity can vary between $1\%$ and $2.5\%$. These values were tested in the simulation.

The reflectivity of the BMV absorber was set to the value of $0.7\%$, as measured in air, Section \ref{Characterization in air}. In LAr this value could change, given the different refractive index of LAr compared to air: values from $0.1\%$ to $0.8\%$ of reflectivity were tried, resulting in a variation of the final results $<1\%$. The thickness of this absorber was set to $120~\mathrm{\mu m}$, according to the material data sheet.

The visible reflectivity of the BAA was investigated, considering that the bottom plate endcap and upper endcap of the sample holder remained uncovered during the measurements. In general, this depends on the type of anodization and surface finish. In the study \cite{Marshall2014} the measured reflectivity at $450~\mathrm{nm}$ of different samples of black-anodized aluminum is $\lesssim10\%$, with fraction of specular component $\lesssim60\%$. We considered these values as a starting point, and then we tuned them as described in the next paragraph.

To estimate the impact of the reflectivity of the stainless steel source disc, different values of the reflectivity were tried in the simulation. Measurements in \cite{Zatschler2020} reported a value of $\sim60\%$ at $450~\mathrm{nm}$. Considering that the external part of the source disc, where the source material is not present, was covered with a piece of black metal velvet absorber, we estimated that only $\sim50\%$ of the source disc was exposed. The resulting reflectivity of the source disc was set to $30\%$. The specular reflectivity of the electropolished stainless steel was measured to be a fraction of $85\%$ of the total reflectivity \cite{Zatschler2020}: this value was considered as an upper limit on the specular reflectivity of the source disc, which is made of unpolished stainless steel. We found negligible impact on the results by varying the disc reflectivity around $30\%$ and the specular component between $40\%$ and $85\%$. 

\subsection{Determination of simulation input parameters with absorber measurements}
\label{abs_sim}

The average number of PE collected, $N_{PE}$, when the effective blue photon yield is $N_{eff}$ and the PMT QE at $450~\mathrm{nm}$ is $QE_{blue}$, is given by
\begin{equation}
    N_{PE} = N_{eff} \cdot QE_{blue}
\label{eq:1}
\end{equation}

The parameter $N_{eff}$ is the mean number of photons emitted by the LED\footnote{This quantity does not represent the physical number of photons emitted per impulse by the LED: it corresponds to a normalization of the simulation, calibrated based on the blue-light BMV measurement.}, $N_{p}$, times a term, $T(\Omega)$, with $\Omega$ being the solid angle subtended by the PMT photocathode, representing the optical transport function, taking into account photon absorption in LAr, Rayleigh scattering, absorptions and reflections by the interfaces, and it is in general function of the photon direction of emission
\begin{equation}
    N_{eff} = N_p \cdot\int  T(\Omega) \frac{d\Omega}{4\pi}
\label{eq:2}
\end{equation}

The blue-light BMV measurement was simulated with $2.76$\,eV photons ($450~\mathrm{nm}$) emitted into the upper hemisphere, toward the sample-holder interior. The value of $N_p$ was adjusted until the simulated mean PE yield matched the BMV measurement in Table~\ref{tab:LArS_results_blue}. An uncertainty of $2\%$ was assigned to $N_p$ based on the BMV measurement.

The BAA measurement exposes a larger area of the holder to blue light and is therefore more sensitive to its optical response. With $N_p$ fixed, the BAA reflectivity parameters were varied to reproduce the measured mean PE yield. The selected total reflectivity was $15\%$, with a specular component of $(75\pm5)\%$. No uncertainty was assigned to the total reflectivity, since its variation was found to have negligible impact on the results.

A total uncertainty of $7.4\%$ was assigned to $N_{eff}$. This was estimated by varying the values of $N_p$ (relative uncertainty contribution to $N_{eff}$ of $2\%$), the BAA specular component ($6\%$), the visible-light absorption length in LAr ($1.5\%$), and the PMT window reflectivity ($3.5\%$) within their respective uncertainties. The individual contributions were then combined in quadrature.

\subsection{Determination of samples blue-light reflectivity in LAr}

\begin{figure}[htbp]
    \centering
    
    \begin{subfigure}{0.48\textwidth}
        \centering
        \includegraphics[width=0.73\linewidth]{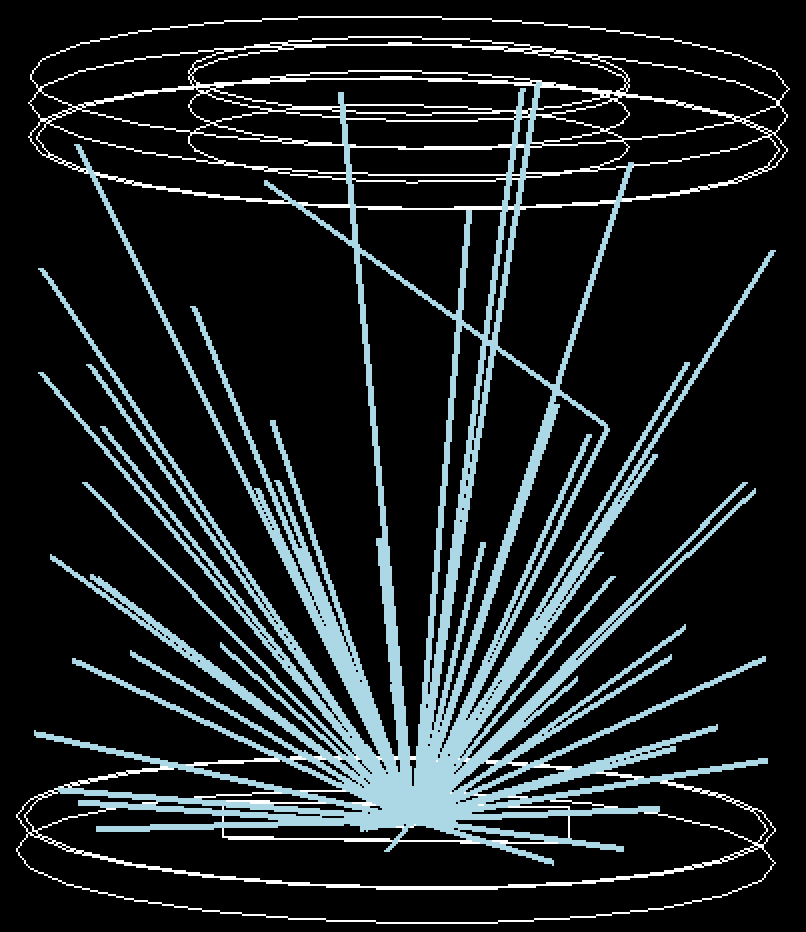}
    \end{subfigure}
    \hfill
    \begin{subfigure}{0.48\textwidth}
        \centering
        \includegraphics[width=0.74\linewidth]{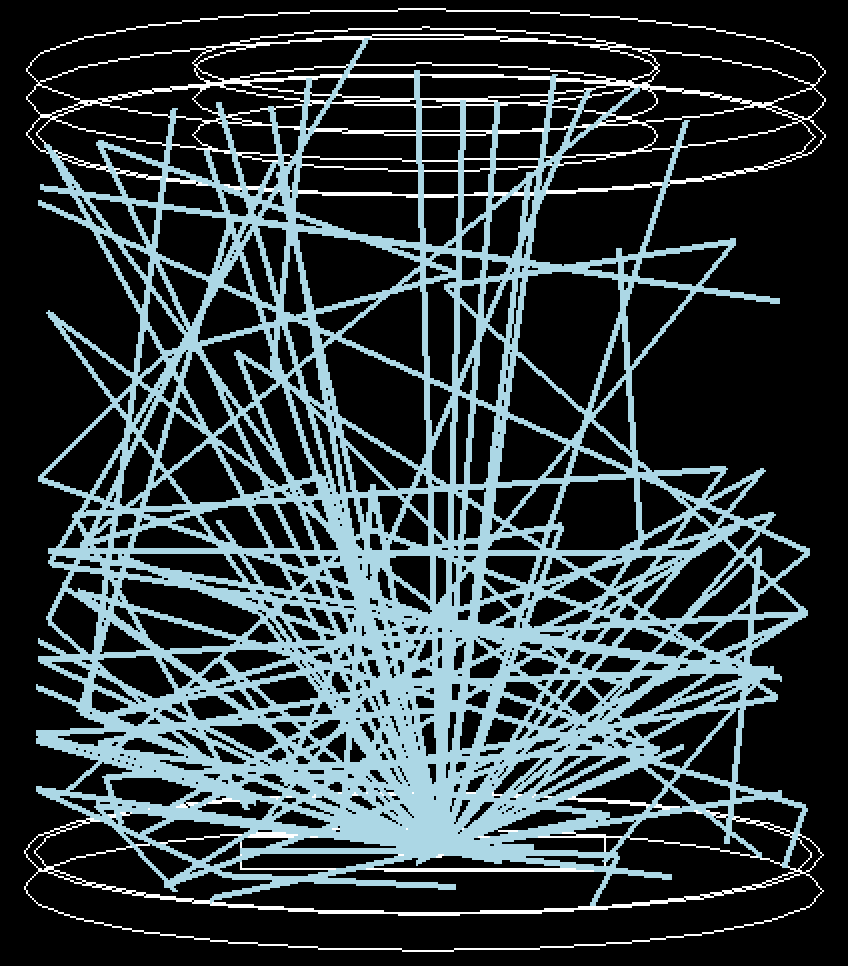}
    \end{subfigure}
    
    \caption{Renderings of two LArS simulations with optical photons trajectories visible, with the BMV absorber on the left image, and the PET reflector on the right. Both renderings are produced using only $100$ photons, for clarity, instead of the full number used in the actual simulations.}
    \label{fig:SimulationFigure}
\end{figure}

With the setup parameters fixed, the PET and 3D-filament printed crystallized PEN reflectivities at $450~\mathrm{nm}$ were adjusted to reproduce the measured PE yields. Figure~\ref{fig:SimulationFigure} illustrates photon transport with an absorber and a reflector. Sample thicknesses were included in both cases. The initial reflectivity values were taken from the air measurements of PET and the crystallized PEN comparison sample (Section~\ref{Characterization in air}), then varied until the simulated and measured mean PE yields matched.

The reflective surfaces were modeled with predominantly diffuse reflection and a small specular component, motivated by the air measurements (Figure~\ref{fig:air_measurements}, right). Varying the specular component between $1\%$ and $5\%$ had negligible impact on the results. Simulations at the central value and lower and upper bounds of $N_{eff}$ were used to estimate the systematic uncertainty. The TTX measurement was simulated with the same approach for comparison. Table~\ref{tab:sim_results} summarizes the inferred reflectivities.

\begin{table}[htbp]
    \centering
    \caption{Reflectivity at $450~\mathrm{nm}$ in LAr of the samples tested.}
    \label{tab:sim_results}
    \smallskip

    \begin{tabular}{c|c}
        \hline
        \textbf{Sample} &
\textbf{Blue light reflectivity in LAr} \\
        \hline
        PET & $(90\pm4)\%$ \\
     
        Crystallized PEN & $(81\pm5)\%$ \\

        TTX & $(80\pm5)\%$ \\
        \hline
        
    \end{tabular}
\end{table}

A VUV simulation of the PET sample coupled to the PEN foil provided a consistency check of the result with the PET sample following the approach of Ref.~\cite{GabrielaArticle}. The VUV optical parameters of LAr, the stainless steel source disc, and PEN, together with their uncertainties, were taken from that work. The simulation was calibrated to the BMV PE yield in Table~\ref{tab:LArS_results_VUV}. The BAA VUV reflectivity was set equal to that of BMV because their measured PE yields were compatible. Using the inferred blue-light reflectivity of PET in Table~\ref{tab:sim_results}, a PEN QE of $69\%$ and the emission spectrum of a PEN foil from Ref.~\cite{GabrielaArticle}, the simulated PE yield agreed with the measurements. This comparison could not be extended to crystallized PEN, as the corresponding emission spectrum and QE were not measured, and therefore could not be assumed to be the same as those of non-crystallized PEN.

\section{Discussion}
\label{sec:discussion}

The measurements demonstrate the potential of microcellular PET and crystallized PEN as alternative reflective materials for LAr-based detectors. At room temperature, PET exhibits a total reflectance approaching $98\%$ in the visible range, with a predominantly diffuse component. In LAr, PET yielded approximately $26\%$ more detected photoelectrons than TTX under blue-light illumination. This enhancement indicates improved light collection in the investigated geometry, although the measured response also depends on the angular distribution of reflected photons and the optical properties of the setup.

The Geant4-based analysis yielded estimated blue-light reflectivities in LAr of $(90\pm4)\%$ for PET, $(81\pm5)\%$ for crystallized PEN, and $(80\pm5)\%$ for TTX. These model-dependent estimates suggest that PET may offer improved reflectivity compared to TTX, whereas crystallized PEN exhibits comparable performance. The extracted values depend on the assumed optical surface properties and photon-transport model, which should be considered when interpreting differences between materials.

Under illumination by LAr scintillation light, PET coupled to a PEN foil and crystallized PEN yielded $(2.36\pm0.11)$ and $(1.75\pm0.10)$ times the detected light of the BMV absorber, respectively. These results demonstrate the effectiveness of the investigated PEN-based configurations in enhancing scintillation-light collection. The uncoated PET sample showed a smaller enhancement of $(20\pm10)\%$, which may arise from VUV reflection, intrinsic luminescence, as previously reported for PET \cite{Ouchi2004FluorescencePET}, or a combination of both. Since the PMT is sensitive to both VUV and visible photons, the present measurements cannot distinguish between these contributions. 

Previous measurements of TPB-coated TTX and PEN foil coupled to TTX provide a useful reference \cite{GabrielaArticle}. However, differences in sample geometry, wavelength-shifting layer thickness, and experimental conditions prevent a direct quantitative comparison. In addition, the triplet lifetimes measured during the present campaign indicate differences in LAr purity between some sample measurements, which may affect the observed scintillation yields.

Under VUV illumination, BAA and BMV gave compatible PE yields in this geometry (Table~\ref{tab:LArS_results_VUV}). Under blue-light illumination, BAA produced a larger yield. The simulation reproduced this response using a total BAA reflectivity of approximately $15\%$ and a specular fraction of approximately $75\%$. These are model-dependent parameter choices; the response of black-anodized aluminum also depends on anodization and surface finish.

The estimated blue-light reflectivity of PET in LAr is lower than its measured reflectance in air. Changes in the refractive-index contrast at material interfaces, together with the microcellular structure of PET, may contribute to this difference. In particular, penetration of LAr into the microcells could modify the scattering properties of the material. Further measurements are required to investigate this possibility and assess the stability of the optical response over longer periods.

For LEGEND-1000, PET coupled to a wavelength shifter offers a highly reflective, predominantly diffuse substrate, whereas crystallized PEN provides wavelength-shifting and reflective properties in a single material. The latter approach could simplify the reflector assembly by reducing the number of material interfaces. However, the optical performance demonstrated here represents only one aspect of material suitability. Radiopurity, cryogenic stability, and performance in larger detector geometries remain to be evaluated.

\section{Summary and outlook}
\label{Summary and outlook}

We characterized the optical response of microcellular PET and crystallized PEN to blue and VUV scintillation light in LAr, complemented by room-temperature reflectance measurements and Geant4-based optical simulations. PET exhibited high diffuse reflectance and enhanced blue-light collection relative to TTX, while the PEN-based configurations substantially increased the detected scintillation-light yield. The simulations yielded blue-light reflectivity in LAr estimates of $(90\pm4)\%$ for PET and $(81\pm5)\%$ for crystallized PEN, within the investigated optical model.

These findings support further investigation of both materials for wavelength-shifting reflector applications in LEGEND-1000 and other LAr-based detectors. Future studies should focus on the optical stability of the materials during extended LAr exposure, the wavelength-shifting properties of crystallized PEN, and direct comparisons with established reflector configurations under matched conditions. Radiopurity screening and tests with larger samples will also be required to assess their suitability for detector-scale implementation.

\appendix

\section{LArS programmable logic controller}
\label{appendix_PLC}

\begin{figure}[htbp]
    \centering
    \includegraphics[width=0.8\textwidth]{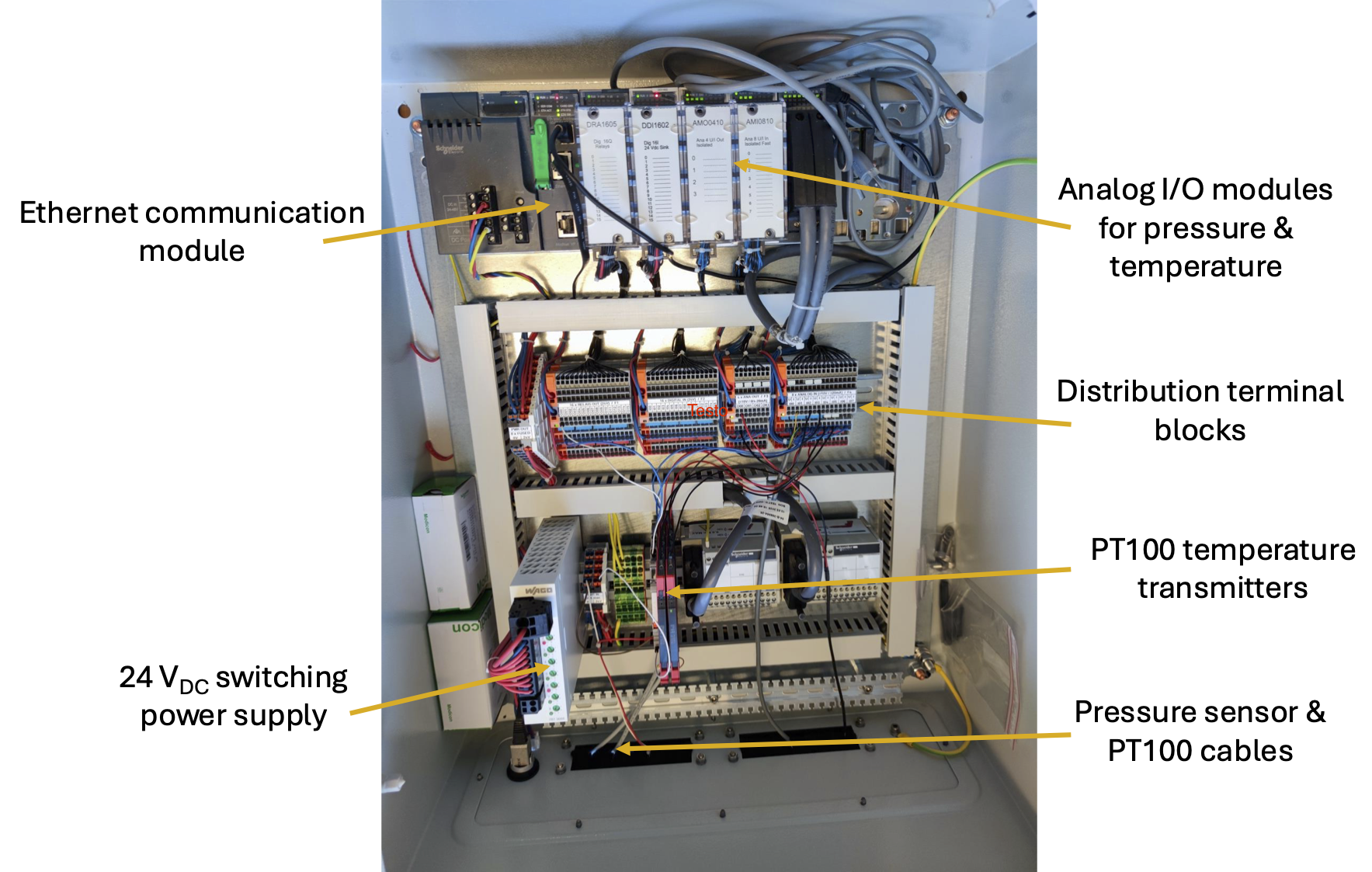}
    \caption{PLC assembled and used for the measurements with LArS.}
    \label{fig:PLC_hardware}
\end{figure}

A programmable logic controller (PLC), shown in Figure~\ref{fig:PLC_hardware}, was assembled to control argon cooling and stabilize the system. An incremental proportional-integral-derivative (PID) algorithm, implemented in EcoStruxure Control Expert, uses the measured pressure as input and adjusts the cooling valve to reach the pressure setpoint. Anti-windup clamping limits delays and pressure overshoots. Temperature, pressure, LAr level, and valve opening are transmitted via Ethernet and displayed in Grafana \cite{grafana}. During the measurements, pressure remained within $\pm30~\mathrm{mbar}$ of the $1.5~\mathrm{bar}$ setpoint, and the measured LAr temperature remained within $\pm0.1~\mathrm{K}$ of $92.6~\mathrm{K}$.

\section{LArS levelmeter}
\label{LArS levelmeter}

\begin{figure}[htbp]
    \centering

    \begin{minipage}[t]{0.53\textwidth}
        \vspace{0pt}
        \centering
        \includegraphics[width=0.6\linewidth]{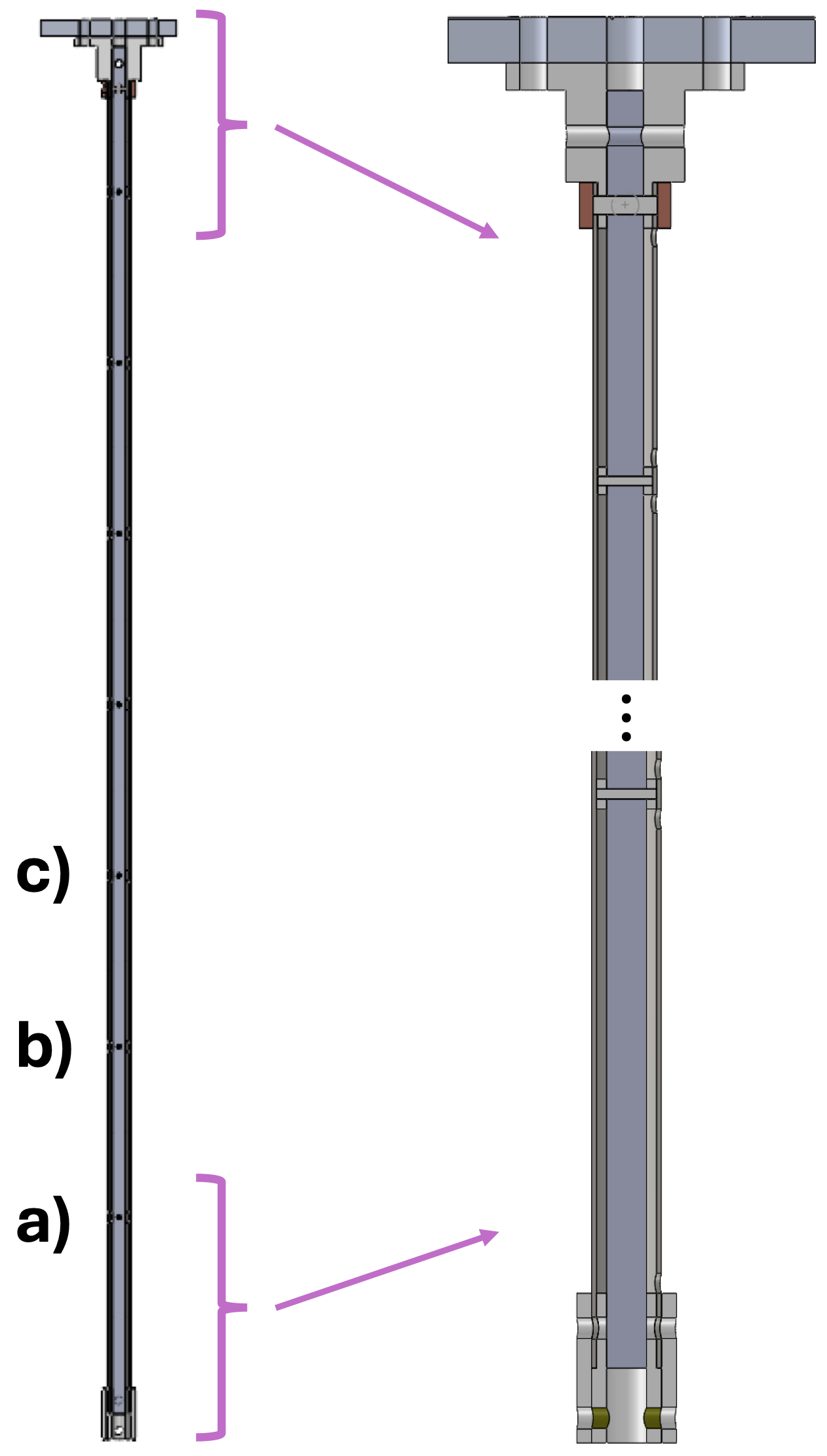}
    \end{minipage}%
    \hfill%
    \begin{minipage}[t]{0.43\textwidth}
        \vspace{0pt}
        \centering
        \includegraphics[width=\linewidth]{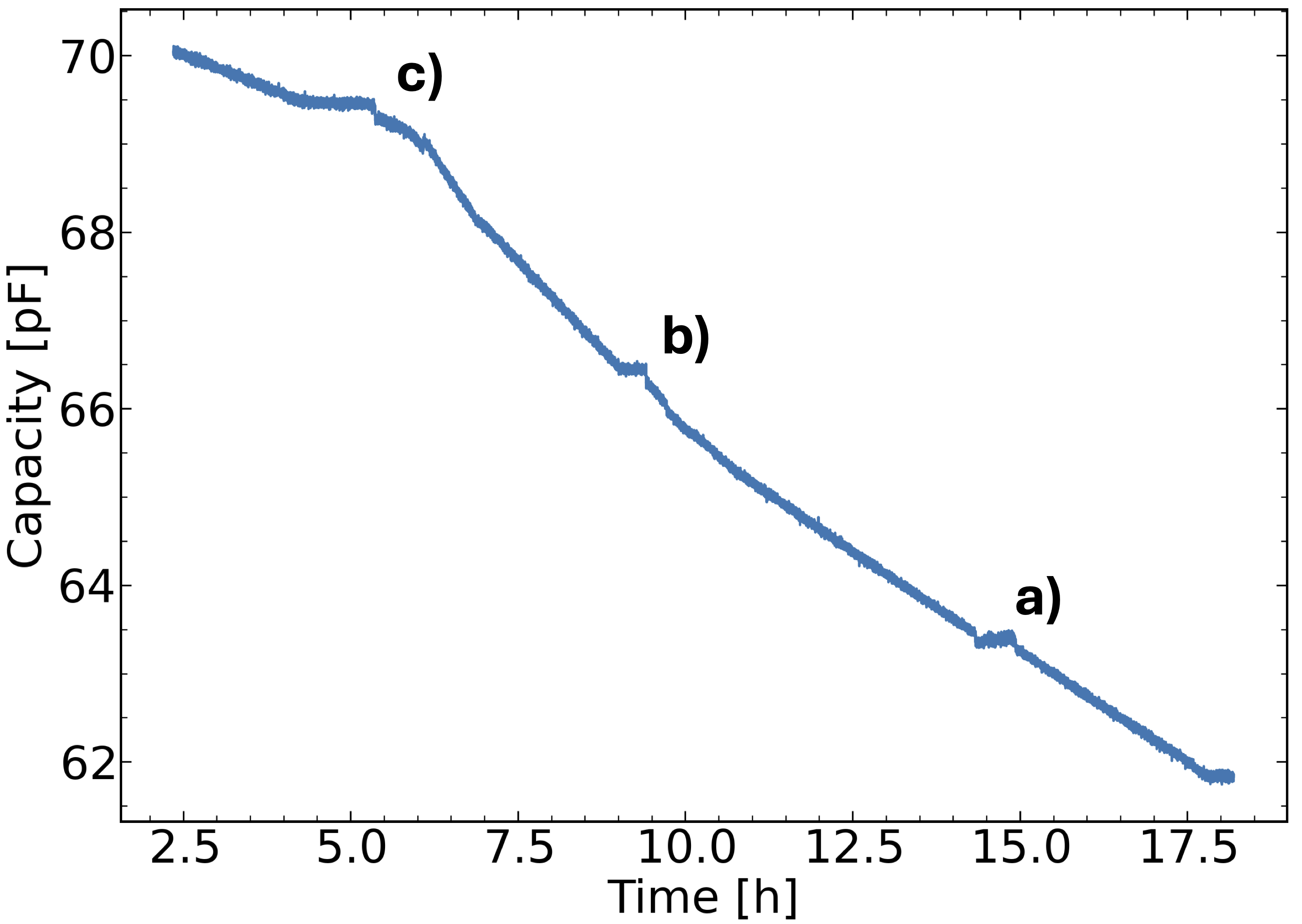}

        \vspace{0.3cm}

        \includegraphics[width=\linewidth]{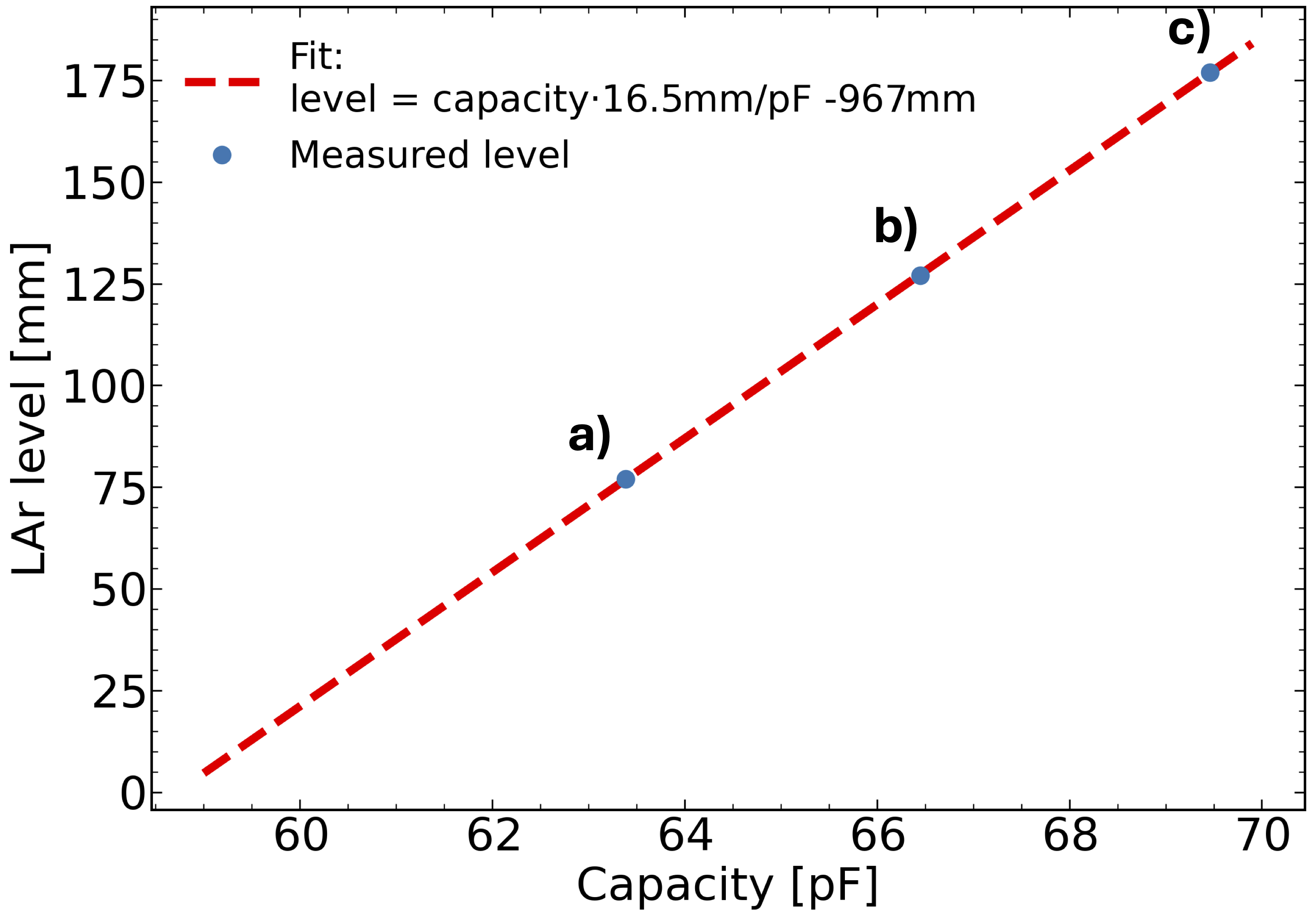}
    \end{minipage}

    \caption{Drawing of LArS levelmeter on the left with insets on the upper and lower parts. Two plots regarding the levelmeter calibration on the right. In the upper plot, the capacity measured at the end of a run during the evaporation of LAr is shown as a function of time elapsed. The three capacity values corresponding to the three levelmeter holes are indicated with a), b), c). In the lower plot, the LAr level is shown as a function of the capacity measured at the three levelmeter holes, together with a fit to the data points.}
    \label{fig:levelmeter}
\end{figure}

The capacitive level meter consists of an inner stainless steel rod and an outer stainless steel tube. Holes spaced by $50~\mathrm{mm}$ along the tube, with PEEK rings at the corresponding positions, provide identifiable levels during filling or evaporation. The known positions and corresponding capacitances are used to calibrate the instrument. Figure~\ref{fig:levelmeter} shows the design and calibration.

\acknowledgments
We thank Cabot-Ann Christofferson and Alissa Love from the South Dakota Mines for cleaning the PEN foil. This research has received funding from the University of Zurich under the UZH Candoc Grant no.~FK-26-096, and from the Swiss National Science Foundation under Grant numbers 2000-1-240028 and 20FL20-232670. This research was also funded by the Deutsche Forschungsgemeinschaft (DFG, German Research Foundation) under Germany's Excellence Strategy – EXC 2094 – 390783311, and by the Deutsche Forschungsgemeinschaft (DFG, German Research Foundation) – SFB 1258 – Project number 283604770. This material is based upon work supported by the U.S. Department of Energy, Office of Science, Office of Nuclear Physics under contract number DE-AC05-00OR22725.

\bibliographystyle{JHEP}
\bibliography{biblio}

\end{document}